\documentclass[twocolumn,iop]{emulateapj}
\usepackage{amsmath,amssymb, xcolor}
\usepackage{graphicx}
\usepackage{natbib}
\usepackage{sistyle}

\usepackage[backref, breaklinks, colorlinks=True,
            citecolor=blue, citebordercolor=red]{hyperref}
\newcommand{\myemail}{vikrant.kamble@utah.edu}
\SIthousandsep{,}

\begin{document}

\title{Measurements of effective optical depth in the L\MakeLowercase{y}$\alpha$ forest \\ from the BOSS DR12 Quasar sample}

\author{Vikrant~Kamble\altaffilmark{1*}, Kyle~Dawson\altaffilmark{1}, H\'elion~du~Mas~des~Bourboux\altaffilmark{1},\\ Julian~Bautista\altaffilmark{2}, Donald~P.~Scheinder\altaffilmark{3,4}}

\altaffiltext{*}{\myemail}

\altaffiltext{1}{Department of Physics and Astronomy, University of Utah, Salt Lake City, UT 84112, USA}

\altaffiltext{2}{Institute of Cosmology and Gravitation, Dennis Sciama Building, University of Portsmouth, Portsmouth, PO1 3FX, UK}

\altaffiltext{3}{Department of Astronomy and Astrophysics, Pennsylvania State
University, University Park, PA 16802, USA}

\altaffiltext{4}{Institute for Gravitation and the Cosmos, The Pennsylvania State University, University Park, PA 16802}

\begin{abstract}
We measure the effective optical depth in the Lyman-alpha (Ly$\alpha$) forest using 40,035 quasar spectra from the Twelfth Data Release (DR12) of the Baryon Oscillation Spectroscopic Survey (BOSS) of SDSS-IV. A rigorous selection based on spectral index and the equivalent width of the C IV emission line is applied to choose seven uniform samples with minimal intrinsic variations across redshifts. Modeling the redshift evolution of the effective optical depth with a power-law, $\tau_\mathrm{eff}=\tau_0 (1 + z)^\gamma$ produces $\tau_0=(5.54 \pm 0.64)\times 10^{-3}$ and $\gamma=3.182 \pm 0.074$. The 2.2\% precision estimate on $\gamma$ is dominated by systematic errors, likely arising from the bias and uncertainties in spectral index estimates. Even after incorporating the systematic errors, this work provides the most precise estimates of optical depth parameters to date. Finally, using the reconstructed Ly$\alpha$ forest continuum to directly measure the transmitted flux ratio as a function of redshift, we find deviations of less than 2.5\% from the predictions from the global model and no convincing evidence for signal associated with He II reionization.

\end{abstract}


\section{Introduction} \label{sec:intro}

Following the formation of the first stars and quasars, the neutral Hydrogen gas present in the Intergalactic Medium (IGM) is ionized in a process known as the epoch of reionization \citep{Zaroubi:2013aa}. The ionization state of the IGM is maintained in equilibrium through a balance of photoionization from background UV radiation and the adiabatic cooling of gas \citep{Haardt:2012aa}. This ionization state evolves with redshift owing to the expansion of the Universe and the change in the rate of photoionization radiation from quasars and massive stars.

The strong absorption features imprinted in the spectra of high-redshift quasars \citep{Gunn:1965aa} provide an important means to study the density, temperature and redshift evolution of the IGM following reionization \citep{Haardt:1996aa, Madau:1999aa}. These absorption features, arising from the resonant scattering of the background light from neutral Hydrogen gas with column densities $N_{\text{HI}} \sim 10^{14} \mathrm{cm}^{-2}$, are known as the Ly$\alpha$ forest \citep{Lynds:1971aa}. Simulations that model the H~I gas following a given temperature-density relation, a given background ionization intensity and perturbations following those of matter have been shown to reproduce the Ly$\alpha$ forest along with various properties remarkably well. One of the most basic quantities that can be obtained from Ly$\alpha$ forest absorption is the effective optical depth of the IGM to Ly$\alpha$ photons. The effective optical depth directly constrains the intensity of the background ionizing flux \citep{McDonald01, Bolton:2005aa} and provides a link between the matter power spectrum and the flux power spectrum measured from Ly$\alpha$ forest \citep{Croft:1998aa, Seljak:2003aa, Tytler:2004aa, Palanque-Delabrouille:2011aa, Chabanier:2018aa}.

Numerous studies using quasar spectra have found a smooth increase in the effective optical depth of the IGM with increasing redshift that can be described by a power-law evolution \citep{Schneider1991}. However, the estimates of the mean opacity at $z = 0$, $\tau_0$, and the exponent of the power-law, $\gamma$, vary between measurements by more than the reported uncertainties. A possible source of these systematic errors is the estimation of the unabsorbed quasar continuum. High-resolution spectroscopic studies perform a direct fitting using the peaks in the Ly$\alpha$ forest to estimate the unabsorbed continuum \citep{Schaye:2003aa, Kim:2007aa, Faucher}. These regions of each spectrum likely lead to underestimates of the continuum level, due to finite optical depth even in the most underdense regions. These systematic underestimates of the continuum level are expected to increase with redshift due to the increase in matter density.

Studies involving large samples of low signal-to-noise ratio ($S/N$) spectra have used composite spectra to obtain high $S/N$ representations of the average absorbed continuum. For this approach to work, it is important that the spectra being averaged have the same underlying unabsorbed continuum over all redshifts. This requirement introduces new complexities to the analysis as one must rely on sample selection based on features at wavelengths greater than the restframe Ly$\alpha$ emission and assume that these trends can be extrapolated into the Ly$\alpha$ forest. Moreover, since one does not know the underlying continuum, this method provides a direct estimate of only the relative optical depth. Previous studies have followed different approaches to model absolute optical depth measurements from low resolution spectra. \cite{Paris:2011aa} used Principal Components Analysis (PCA) to model the continuum at wavelengths longer than Ly$\alpha$ emission to predict the continuum in the Ly$\alpha$ forest. One can also perform a joint modeling of a parameterized continuum and optical depth evolution using the flux in the Ly$\alpha$ forest \citep{Bernardi03, Prochaska:2009aa}. Another approach is to perform measurements of the relative optical depth and to fit those measurements with a model for the evolving optical depth \citep{Becker:2013aa}. Our modeling of the Ly$\alpha$ forest flux measurements is similar to those of  \cite{Becker:2013aa}, except that we do not use composite spectra.

\cite{Bernardi03} found a sharp bump at $z \sim 3.2$ of width $\Delta z \sim 0.4$ in the optical depth evolution. They attribute this feature to the increase in temperature of the IGM following He II reionization. A careful study using high $S/N$, high resolution quasar spectra by \cite{Faucher} found a similar feature. However, other studies have not detected the bump \citep{Kirkman:2005aa, Paris:2011aa}. Recent measurements using composite spectra created from a sample of 6065 quasar spectra \citep{Becker:2013aa} also reveal no such feature.

Given the size of the dataset available from the final Baryon Oscillation Spectroscopic Survey \citep[BOSS;][]{Dawson:2013aa}, we can explore the systematic errors that arise from modeling assumptions while also improving the statistical constraints on the redshift evolution of the mean Ly$\alpha$ transmission.
Following a recent study on quasar spectral diversity by \cite{Jensen:2016aa}, we control the sample to have similar
physical properties by dividing the spectra into seven bins based on observable parameters: spectral index and Carbon IV (C IV) equivalent width.

Assuming a smooth evolution of the effective optical depth modeled as a power-law, we constrain its parameters from each of these seven subsamples. The major differences from previous work are:
\begin{enumerate}
\item We bin by quasar properties and make seven independent measurements of the effective optical depth. This process allows a test of assumptions about uniformity of quasar continuum in the Ly$\alpha$ forest over the full redshift range.
\item We account for correlations between rest-frame wavelengths primarily due to cosmological fluctuations on small scales.
\item We quantify the systematic errors in the analysis by introducing systematic covariance matrices.
\end{enumerate}
These refined measurements can supplement other areas of research that involve Ly$\alpha$ forest such as the measurement of Baryon Acoustic Oscillations (BAO) feature \citep{Bautista:2017aa, du-Mas-des-Bourboux:2017aa} and the one-dimensional power spectrum \citep{Palanque-Delabrouille:2013aa, Chabanier:2018aa}.

The structure of this paper is as follows: we describe our data and sample selection in Section~\ref{sec:data}. We also detail how the observable parameters were estimated and the methods used to correct for flux miscalibrations and measurement uncertainty. Section~\ref{sec:optdepth} describes the method used to obtain optical depth estimates from the raw flux data. Possible sources of systematic errors in our measurements are investigated in Section~\ref{sec:systematics}. The reconstructed continuum for each bin and their interpretation are presented in Section~\ref{sec:interpretation}. Comparison of our results to optical depth measurements from previous studies are presented in Section~\ref{other_results}, along with a discussion on the evidence of a He II feature. We summarize the analysis in Section~\ref{sec:conclusion}.

\section{Data  and sample selection} \label{sec:data}
The quasar spectra used to measure the effective optical depth were obtained from BOSS \citep{Dawson:2013aa}, a part of the third generation of the Sloan Digital Sky Survey \citep[SDSS-III;][]{Eisenstein:2011aa}. The primary goal of BOSS was to extract cosmological constraints from BAO that behave like a standard ruler \citep{ Alam:2017aa, Bautista:2017aa, du-Mas-des-Bourboux:2017aa}. This section describes how we identify an appropriate spectroscopic quasar sample for this study. We then characterize the spectral diversity of the sample and bin by common features. Finally, we perform corrections to the flux calibrations and uncertainties in the flux estimates.

\subsection{Spectroscopic data}
BOSS uses a pair of double spectrographs \citep{Smee:2013aa} mounted on the Apache Point 2.5 m Telescope \citep{Gunn:2006aa}. The quasar selection involves a combination of algorithms \citep{Kirkpatrick:2011aa, Palanque-Delabrouille:2011aa, Bovy:2011aa} that are detailed in \cite{Ross:2012aa}. The observations are conducted using aluminum plates; each plate subtends an angle of $3^o$ on the sky and contains 1000 holes of $2''$ diameter drilled at locations corresponding to spectroscopic targets. Optical fibers are manually inserted into each plate to feed the pair of spectrographs. After observations of roughly one hour per plate, the raw data are processed into one-dimensional spectra and classified \citep{Stoughton:2002aa, Bolton:2012aa}. The reduced spectra used in this analysis correspond to version {\tt v5\_10\_0} of the pipeline presented in the 14th Data Release \citep[DR14;][]{Abolfathi:2018aa}. This version of the data reduction pipeline corrects for Atmospheric Differential Refraction \citep{Margala:2016aa, Jensen:2016aa}. This dataset is the first spectroscopic sample released publicly from eBOSS \citep{Dawson:2016aa}, a component of  SDSS-IV \citep{Blanton:2017aa}.

We use the quasar classifications from the quasar catalog \citep[DR12Q;][]{Paris:2017aa} released with the 12th Data Release \citep[DR12;][]{Alam:2015aa}.  We include the best spectrum of each quasar with \texttt{ZWARNING}=0. The redshift estimates obtained from visual inspection (\texttt{Z\_VI}) were used as the systemic redshift of the quasars. Quasars with broad absorption lines (BALs) were removed using the \texttt{BAL\_FLAG\_VI} attribute specified in DR12Q. Quasars containing absorption from Damped Lyman-Alpha systems (DLAs) identified in the updated catalog from  \cite{Noterdaeme:2012aa} were also excluded. We select quasars with $z_q > 1.6$ and a median $S/N$ per pixel greater than five computed over the bandpasses $1280 < \lambda_\mathrm{rf} < 1290$, $1320 < \lambda_\mathrm{rf} < 1330$, $1345 < \lambda_\mathrm{rf} < 1360$ and $1440 < \lambda_\mathrm{rf} < 1480$ \AA\ in the quasar restframe. This cut on $S/N$ was made to obtain a reasonable precision in the placement of quasars into their appropriate bins based on the observable properties. The sample size after application of each criterion is reported in Table~\ref{tb:criterion}. The $S/N$ cut produces the largest fractional change in the sample size.

To correct for the effects of Galactic extinction, we adopt the Fitzpatrick model \citep{Fitzpatrick:1999aa} and the Galactic dust extinction map from \cite{Schlegel}.
Possible contamination from atmospheric emission were removed using a skyline mask from \cite{Delubac15}. Measured flux at $\lambda_\mathrm{obs} < 3700$ \AA\ was excluded because of lower $S/N$ and increased uncertainty in flux calibration. Each spectrum was shifted into its restframe wavelength solution in logarithmic units with the same pixel width as that of BOSS. The shifts were performed to the nearest pixel to avoid resampling.
% \textcolor{red}{The fractional error between pixels in restframe given the error on redshift using:
% \[\frac{\Delta \lambda_\mathrm{rf}}{\lambda_\mathrm{rf}} = \frac{\delta z_q}{1 + z_q},\]
% is $0.3 \times 10^{-3}$, at a typical $z_q = 2.5$ and a conservative estimate of $\delta z_q = 10^{-3}$. The pixels however are separated by $\delta \log \lambda_\mathrm{rf} = 0.1 \times 10^{-3}$, which are three times smaller in separation compared to the effects arising from redshift errors.}

\begin{deluxetable}{lrc}
\tabletypesize{\scriptsize}
\tablecaption{The Number of Quasars Remaining in the Sample After Each
Selection Criterion is Applied\label{tb:criterion}}
\tablewidth{0pt}
\tablehead{
    \colhead{Selection} & \colhead{Remaining} &  \colhead{Percent} \\
    \colhead{Criterion} & \colhead{quasars} &  \colhead{remaining}
}
\startdata
All & \num{297301} & -- \\
Objects with \texttt{ZWARNING} = 0 & \num{283405} & 95\\
Remove quasars with BALs & \num{257138} & 86\\
Remove quasars with DLAs & \num{231785} & 78\\
$z_q > 1.6$ & \num{163263} & 55\\
$S/N > 5$ & \num{58062} & 19\\
Cuts in parameter space& \num{40035} & 13
\enddata
\end{deluxetable}

\subsection{Sample selection} \label{sec:sample_selection}
It is crucial to minimize spectral diversity in the quasar sample so the redshift evolution of the observed flux levels in the Ly$\alpha$ forest can be attributed to the IGM and not to the quasars themselves. One prominent source of diversity originates from the Baldwin Effect, wherein luminosity is found to be anti-correlated with equivalent widths (EW) of broad emission lines \citep{Baldwin:1977aa}. \citet{Jensen:2016aa} found the spectroscopic signature associated with luminosity to be dominated by emission lines across the spectrum, a trend generally consistent with the Baldwin effect. It is likely that the variations in line strengths are correlated with variations in intrinsic quasar properties rather than simply luminosity. For example, \cite{Richards:2011aa} argue that the C IV emission varies with quasar winds and the spectrum of the ionizing continuum. The color of the quasar defined by the spectral index is another feature of spectral diversity. E.g., \cite{Ivashchenko:2014aa} report variations in spectral index lead to systematic errors in redshift estimates.

We introduce a new technique of measuring optical depth from low resolution spectroscopy compared previous studies \citep{Bernardi03, Becker:2013aa} by exploring spectral diversity in the sample. This refinement is enabled by the sheer number of quasar spectra in the BOSS sample.

Motivated by the observations above, we assume that a significant fraction of the variations in the Ly$\alpha$ forest continuum arise from variations in spectral index ($\alpha_\lambda$) and C IV equivalent width $W_\lambda(\mathrm{C IV})$. Spectral index captures the continuum level and its shape in the Ly$\alpha$ forest relative to the shape of the continuum at wavelengths longer than 1216 \AA. Variations in emission line features in the forest, particularly the strong O VI feature, are likely to be related to variations in C IV equivalent width. Hence to reduce the diversity within each sample, we first measure $\alpha_\lambda$ and $W_\lambda$(C IV) for each quasar so that the quasars can be binned in a grid spanned by those parameters.

\begin{figure}
\centering
\includegraphics[width=0.5\textwidth]{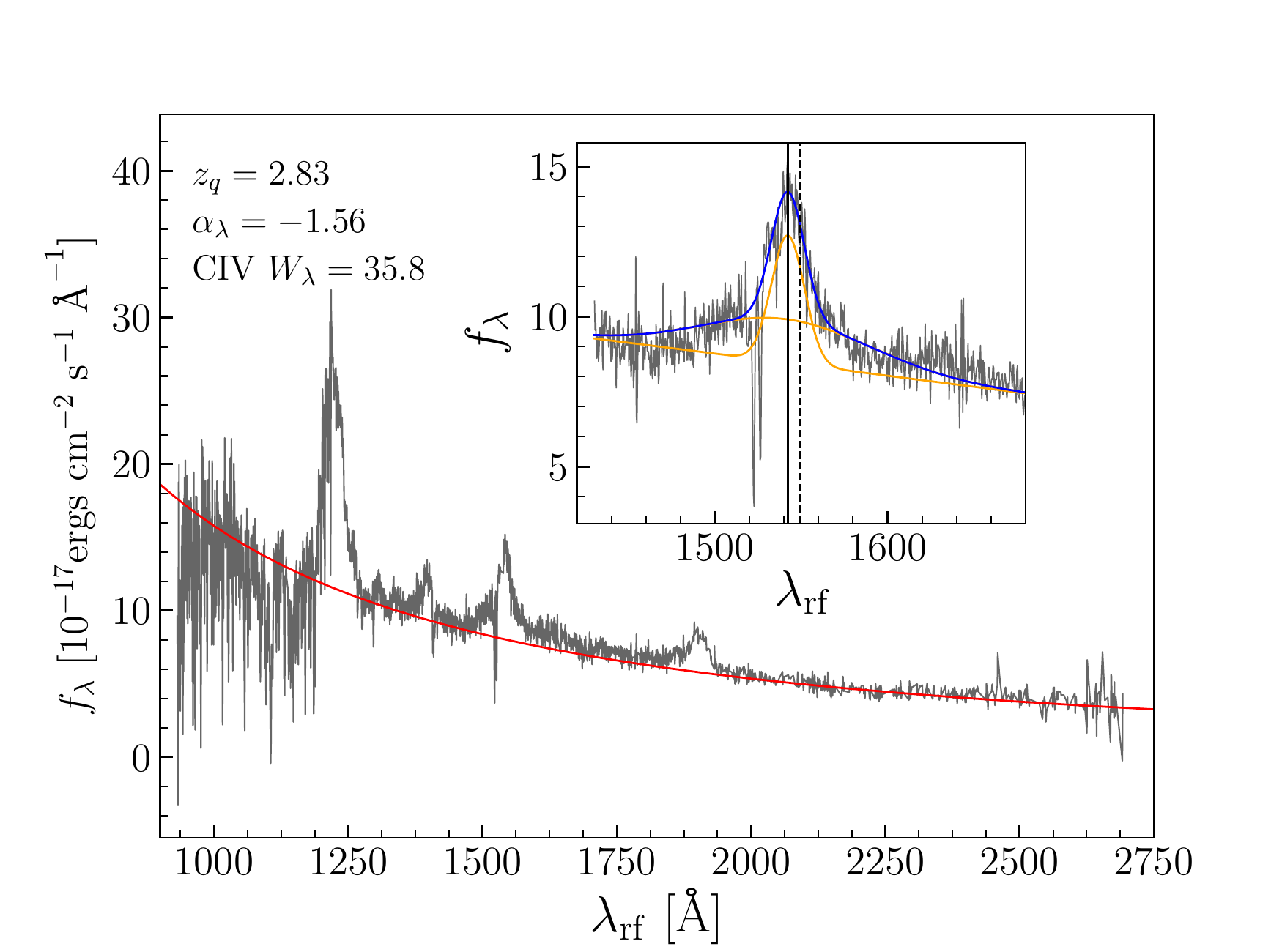}
\caption{A sample quasar at redshift 2.83 with Plate = 6151, MJD = 56265 and Fiber=88. The pseudo-continuum fit is parameterized by a power-law (red), while the C IV emission (shown in inset) is fit by a double Gaussian (blue).}
\label{fig:demo}
\end{figure}

To estimate the spectral index, $\alpha_\lambda$, a pseudo-continuum of the form $f = b\ \lambda^{\alpha_\lambda}$ is fit to each spectrum over the range $1280-1290, 1320-1330, 1345-1360$ and $1440-1480$~\AA. These regions are relatively free of emission lines after visually inspecting the high $S/N$ composite spectrum from \cite{Harris:2016aa}. To ensure reliable estimates of spectral index, only spectra that contain more than 20 good pixels in each of the wavelength intervals were used. The fit was iterated three times, each time clipping points that were three standard deviations below the median estimate. This outlier rejection was performed to mitigate the bias arising from narrow metal absorption lines. We do not use restframe windows above C IV (1550~\AA)\ emission to avoid contamination by iron emission line complexes. Unfortunately, the values of the spectral indices depend strongly on the restframe ranges used for the calculation. However, since our aim is to create a uniform sample across redshift for each bin, potential error in spectral index determination should not bias the results as long as the measurements do not vary with redshift. We explore this assumption further in Section~\ref{sec:systematics}.

For calculating the equivalent width of C IV emission, the line is modeled as a sum of two Gaussians. A simple estimation of the underlying continuum was performed using a linear fit over the range $1450-1465$~\AA\ and $1685-1700$~\AA. The continuum-subtracted spectrum was fit by a double Gaussian over the wavelength range $1500-1580$~\AA. It is widely known that the broad and narrow components of the line arise from gas with different kinematics \citep{Marziani:2010aa}; hence the parameters (location, scale and amplitude) of the two components are varied independently. We excluded pixels at wavelengths larger than 1580 \AA\ to avoid contamination from He II (1640 \AA) emission. A first stage of rejection of possible absorption lines was performed by smoothing with a box kernel of size 20 pixels and then removing pixels from the raw spectrum whose flux values were three standard deviations below the smoothed spectrum. Iterative clipping against the two component Gaussian model was then performed, rejecting negative outliers at the three standard deviation level.

\begin{figure}
\centering
\includegraphics[width=0.5\textwidth]{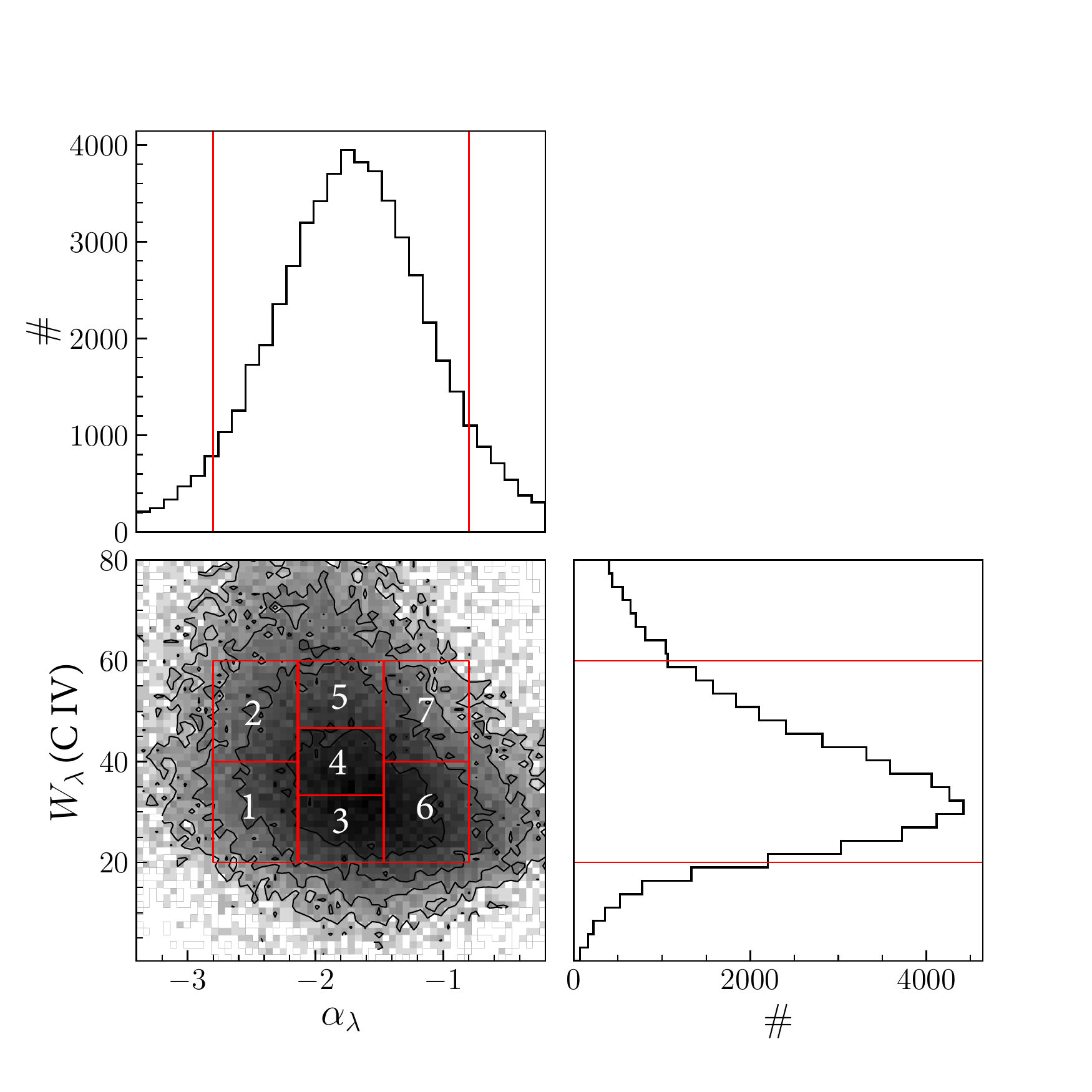}
\includegraphics[width=0.5\textwidth]{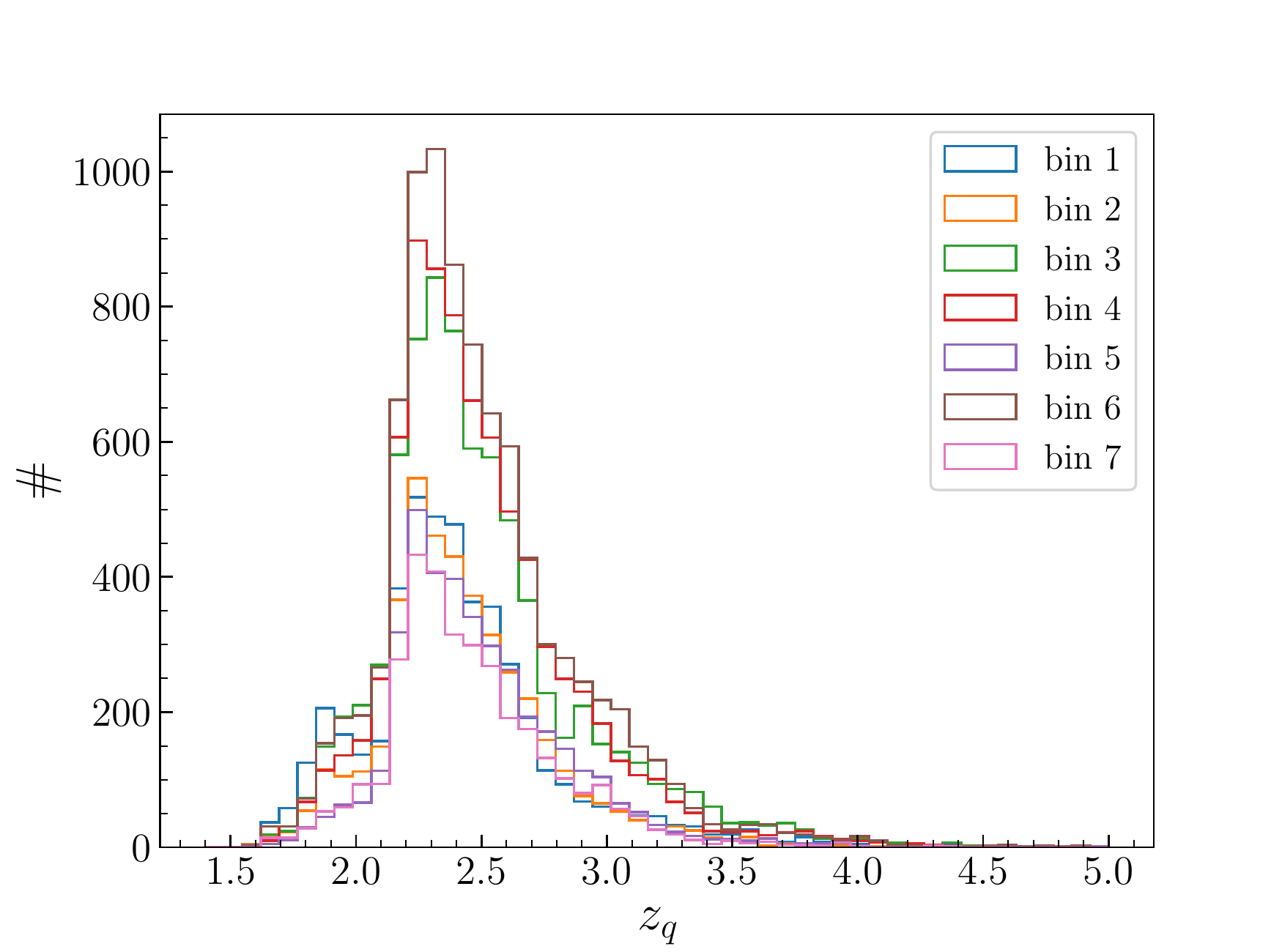}
\caption{{\bf Top:} Distribution of the 58,062 quasars that satisfied our $S/N$ criterion over the
parameter space of $\alpha_\lambda -$ $W_\lambda$(C IV), with the nomenclature for the seven bins. {\bf Bottom:} Distribution of quasar redshifts for each of the seven bins.}
\label{fig:sample_dist}
\end{figure}

Figure~\ref{fig:demo} shows the fit of power-law continuum and two component Gaussian for a sample quasar. The distribution of all 58,062 quasars that meet our $S/N$ criterion in the $\alpha_\lambda -$ $W_\lambda$(C IV) plane is shown in the top panel of Figure~\ref{fig:sample_dist}.

To remove the contribution from varying luminosity, the spectra were normalized using the median value calculated in the restframe wavelength window $1450-1460$~\AA. The objects were then divided into seven bins in the $\alpha_\lambda - W_\lambda(\mathrm{C IV})$ plane. The bins were created using the central 84\% of objects in the projected $\alpha_\lambda$ distribution and the central 84\% of objects in the projected $W_\lambda$(C IV) distribution. Three bins spaced evenly in $\alpha_\lambda$ were created from the remaining sample. The first and third $\alpha_\lambda$ bin were divided in half according to $W_\lambda$(C IV), while the middle bin was divided into three equal parts. This binning scheme and the numbers assigned to each bin are illustrated in the top panel of Figure~\ref{fig:sample_dist}. The bottom panel of Figure~\ref{fig:sample_dist} presents the redshift distribution of quasars for each bin. The seven bins contain a total of 40,035 quasars.

Objects in each bin are then warped by a power-law to reduce broadband continuum variations that may appear in the Ly$\alpha$ forest. The common spectral index to which they are warped was chosen to be the mean spectral index of all the objects in a given bin.

We then require that the $\alpha_\lambda -$ $W_\lambda$(C IV) distribution of objects be identical across quasar redshifts. For this purpose, we first discretized the quasar redshifts into intervals of width 0.2, with the first interval being $2.1 < z_q < 2.3$. Weights are assigned to each object as a function of $\alpha_\lambda$ and $W_\lambda$(C~IV), such that the weighted probability distribution is identical across all redshift intervals. These weights are multiplied by the inverse variance vector for each object, thus reducing biases owing to selection effects. We test our assumptions of mitigating spectral diversity in each bin in Section~\ref{sec:interpretation}.

\subsection{Calibration} \label{sec:calibration}
\begin{figure}
\centering
\includegraphics[width=0.5\textwidth]{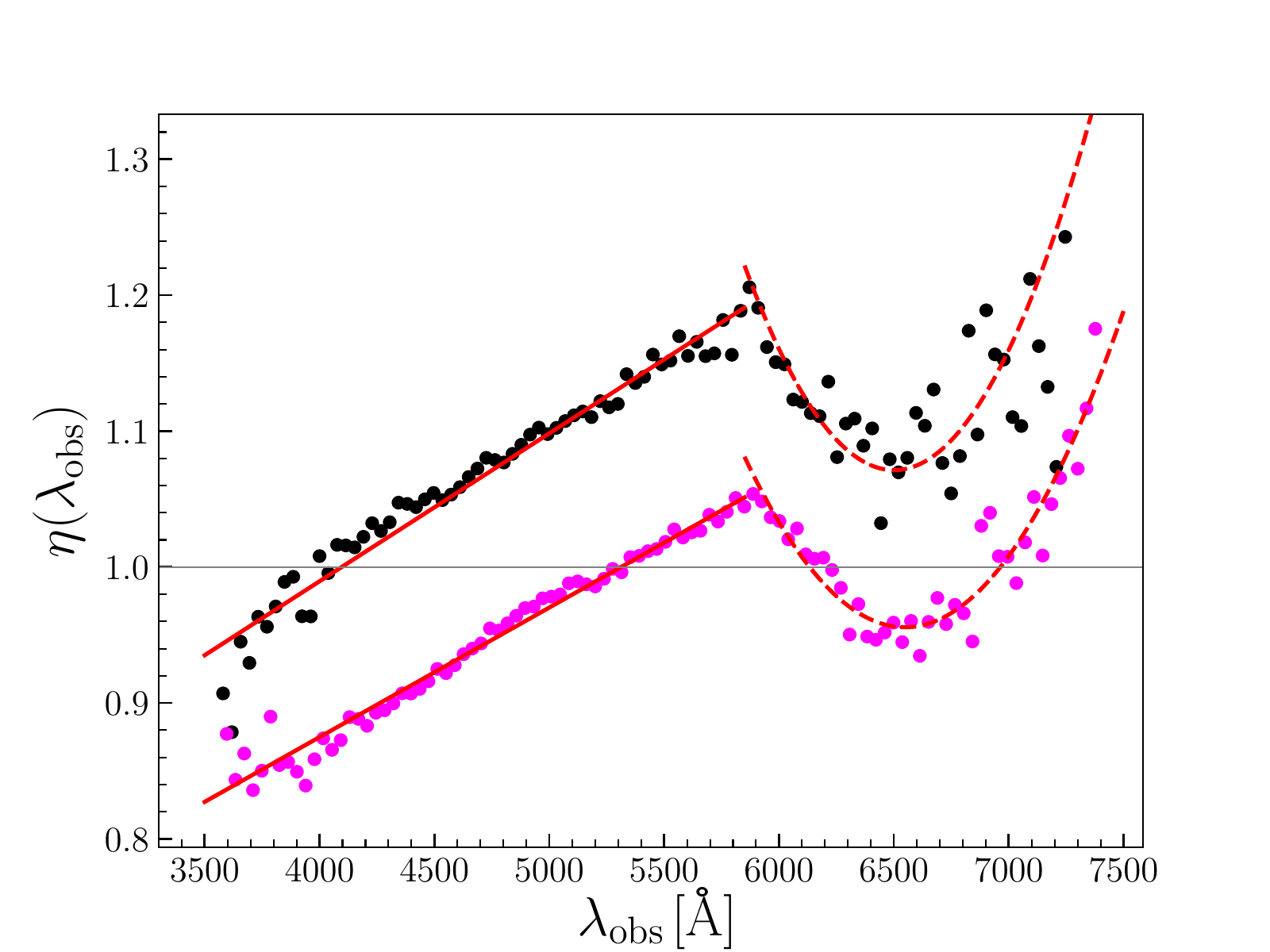}
\includegraphics[width=0.5\textwidth]{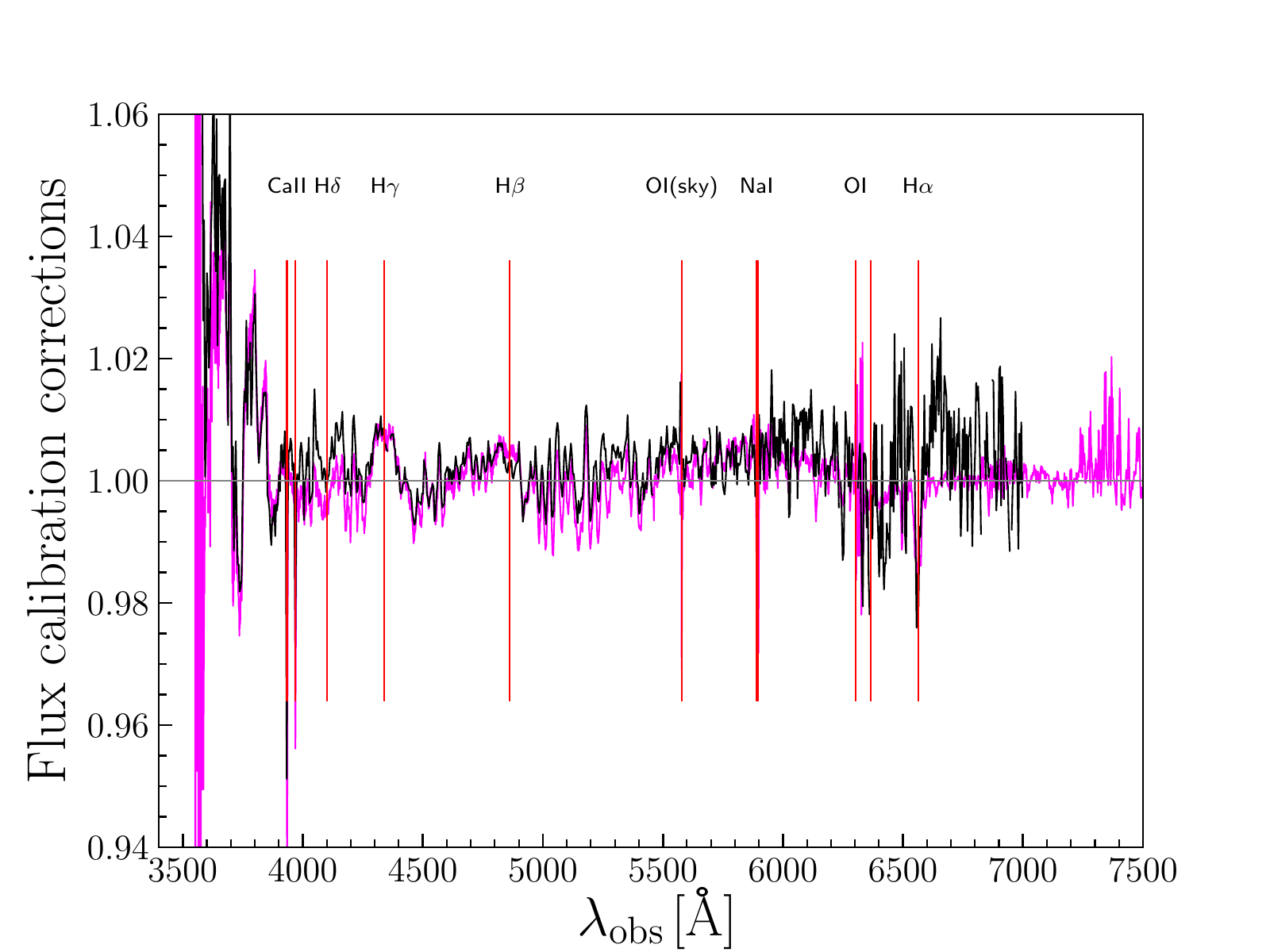}
\caption{{\bf Top:} Ratio of pipeline noise estimates to the flux dispersion in quasars with $1.6 < z_q < 4.0$ using a 10 \AA\ restframe bandpass ($black$) and a 50 \AA\ bandpass ($magenta$). The pipeline is found to overestimate the error by as high as 20\%. A linear fit over the wavelength range $3600-5800$ \AA\ and a quadratic fit over the range $5800-7400$ \AA\ are shown in red. {\bf Bottom:} Flux calibration corrections obtained by stacking residuals as a function of observed wavelength (black). Shown in magenta are the corrections as found by \cite{Bautista:2017aa}.}
\label{fig:calib}
\end{figure}

BOSS spectra have been shown to suffer from systematic errors in the estimates of measurement uncertainty assigned by the reduction pipeline \citep{Palanque-Delabrouille:2013aa, Delubac15}. To recalibrate the uncertainty estimates, we follow a similar analysis as was done in \cite{Lee:2013aa}. We identify a bandpass of width 10 \AA\ in the restframe with relatively few emission lines or gradient in the continuum level. For each quasar spectrum, we perform a $\chi^2$ fit against an average flux level and rescale the pipeline uncertainty until we achieve a $\chi^2$ equal to the number of degrees-of-freedom. The sideband ranges chosen for this process are $1350-1360$ \AA\ and $1470-1480$ \AA. This ratio was averaged in the observer frame over all quasars with $1.6 < z < 4.0$. The fractional error in the pipeline uncertainty estimate, $\eta(\lambda_\mathrm{obs})$, as a function of observed wavelength, is displayed in the top panel of Figure~\ref{fig:calib}. The pipeline overestimates the measurement error for $\lambda_\mathrm{obs} > 4000$~\AA\, with a maximum difference of $20 \%$ at $ \lambda_{\mathrm{obs}} \approx 5800$~\AA. We find an overall shape that is roughly consistent with that measured by \citep{Palanque-Delabrouille:2013aa} but with a positive offset. The difference could be caused by the new reduction algorithms that were introduced in DR14. Earlier studies also used a larger sideband range of 50 \AA, potentially leading to additional dispersion caused by spectral features in the quasars. To examine this assumption, we repeat our analysis using the same 50 \AA\ sidebands and find a suppression in $\eta(\lambda_\mathrm{obs})$, similar to previous studies. In this work, we apply corrections to the estimates of the flux uncertainty using the 10 \AA\ sideband results.

Flux calibration in BOSS relies on theoretical models for F-type stars used as spectroscopic standards. Incorrect modeling of stellar features or Galactic absorption can distort estimates of the Ly$\alpha$ transmission in a redshift-dependent fashion. We assume that all quasars in a bin share the same spectral properties at wavelengths longer than 1216 \AA\ and measure variations in the restframe spectrum as a function of observed wavelength. We attribute systematic difference at any observed wavelength to flux calibration errors. In correcting these errors, we use the same ranges that were employed to measure the spectral indices. We model the flux as a function of observed wavelength over each restframe wavelength pixel, normalized with the average flux measured over the observer wavelength range $4600-4640$ \AA. The results are presented in the bottom panel of Figure~\ref{fig:calib}. The flux calibration errors are in agreement with those presented in \citep{Bautista:2017aa, Lan:2018aa}. The flux calibration appears to deviate by few percent at wavelengths below $3700$ \AA. In the subsequent analysis, we only use flux measurements at observed wavelengths greater than this value. It is important to note that we cannot remove large scale flux calibration errors, as they are degenerate with the power-law distortion applied to each quasar spectrum. We revisit the effects of absolute flux calibrations on the estimation of optical depth parameters in Section~\ref{sec:systematics}.

While we apply these corrections to measurement uncertainties and flux calibration, we are confident that they have no significant effect on the analysis. Hence we do not investigate their effects in detail.

\section{Optical depth measurements} \label{sec:optdepth}
Optical depth studies have historically used high-resolution spectra with direct models for the continuum or relied on composite spectra built from large samples of low-resolution data \citep[e.g.][]{Bernardi03, Becker:2013aa}. We constructed a large sample of low-resolution spectra but make use of the raw flux values at each pixel for each quasar rather than composite spectra. This allows one to model the variance contribution from Large Scale Structure into the likelihood definition. With the seven quasar bins available, we perform seven independent measurements of the redshift evolution of optical depth. Since the quasars in each bin illuminate the same global matter density field, any differences in the derived optical depth can be attributed to systematic errors associated with residual quasar diversity or with the parameter estimation.

\subsection{General framework} \label{sec:framework}
The Ly$\alpha$ forest region used for this analysis covers the restframe wavelength range $1070 \le \lambda_{\mathrm{rf}} \le 1160$ \AA. In this region of each quasar spectrum, the average observed flux, $\bar{f}(z, \lambda_{\mathrm{rf}})$, at a given pixel is given by:
% in restframe wavelength $\lambda_{\mathrm{rf}}$ and redshift $z$, is given by:
\begin{equation} \label{eq:transmission}
\bar{f}(z, \lambda_{\mathrm{rf}}) = C(\lambda_{\mathrm{rf}})\ \overline{T(z)},
\end{equation}
where $C(\lambda_{\mathrm{rf}})$ represents the unabsorbed quasar continuum and $\overline{T(z)}$ is the mean transmission of the neutral Hydrogen in the IGM that gives rise to the Ly$\alpha$ forest. The mean transmission is related to the effective optical depth according to the relation $\overline{T(z)} = e^{-\tau_\mathrm{eff}(z)}$. Since the Ly$\alpha$ absorption takes place at the restframe of the absorber, the absorber redshift $z$ is related to the observed wavelength $\lambda_\mathrm{obs}$ as:

\begin{equation} 1 + z = \frac{\lambda_{\mathrm{obs}}}{1215.67\ \mathrm{\AA}} .\end{equation}
However, since we are working in the restframe of the quasar, the restframe wavelength $\lambda_\mathrm{rf}$ is also related to the observed wavelength as:
\begin{equation}
\lambda_\mathrm{rf}(1 + z_q) = \lambda_\mathrm{obs},
\end{equation}
where $z_q$ is the quasar redshift.

The mean transmission is the product of Ly$\alpha$ and metal absorptions at different redshifts whose contributions are degenerate with those from neutral Hydrogen. In Subsection~\ref{sec:metals}, we argue that metals do not significantly affect the measured redshift evolution of neutral Hydrogen density, but primarily change the estimation of the quasar continuum.

As in previous studies \citep{Press:1993aa, Bernardi03, Kim:2007aa}, we parameterize the effective optical depth by a power-law:
\begin{equation}
\tau_{\mathrm{eff}}(z) = \tau_0 (1 + z)^\gamma.
\end{equation}
Each flux value, $f(z, \lambda_{\mathrm{rf}})$, from each quasar, is assumed to be drawn from a Gaussian distribution with mean $\bar{f}(z, \lambda_{\mathrm{rf}})$ and variance $\sigma^2 = \sigma_f(z)^2 + e_i^2$. The measurement error, $e_i$, is assigned after taking into consideration read noise, photon noise, and other processes and corrected as per Subsection~\ref{sec:calibration}. The redshift-dependent contribution from large-scale structure (LSS), $\sigma_f(z)$, is approximated to be Gaussian and hence added in quadrature to the measurement errors.

 Following \cite{Lee:2015aa}, the variance in the transmission from Large Scale Structure is modeled as:
\begin{equation}
\sigma^2_T(z) = A \left(\frac{1+z}{1+z_r}\right)^B \overline{T(z)}^2,
\end{equation}
where $z_r = 2.25$, $ A = 0.065$, and $B = 3.8$ as given in \cite{McDonald:2000aa}.
However, since we are using measurements of flux, the variance in transmission is translated to variance in flux according to the relation:

\begin{equation}\sigma^2_f(z, \lambda_\text{rf}) = A \left(\frac{1+z}{1+z_r}\right)^B \bar{f}^2(z, \lambda_\text{rf}).\end{equation}

A joint likelihood fit was performed for all quasars in a given bin at a given restframe wavelength. Our model thus consists of three free parameters: the unabsorbed continuum $C(\lambda_\text{rf})$ and the two optical depth parameters $\tau_{0}$ and $\gamma$. The posterior distribution was marginalized over the parameter $C(\lambda_\mathrm{rf})$  to create the likelihood contours over $\tau_0$ and $\gamma$. This process was performed over all the restframe wavelength pixels in the forest range $1070 \le \lambda_\mathrm{rf} \le 1160\ \mathrm{\AA}$. We use a more conservative range than BOSS BAO studies \citep{Bautista:2017aa, du-Mas-des-Bourboux:2017aa} to mitigate contamination from Ly$\alpha$ emission, O VI emission, and extrapolation of the power-law correction to short wavelengths. We also performed fitting allowing $A$ and $B$ to be free parameters. The best-fit values were consistent with \cite{McDonald:2000aa} and did not significantly change the $\tau_{0} - \gamma$ contour, hence they were fixed as described above.

The final likelihood surface over the optical depth parameters for each bin of quasars was obtained by combining the contours over all 351 restframe wavelengths.
Since there is a strong correlation between $\ln \tau_0$ and $\gamma$, it was computationally efficient to sum the likelihood in a rotated basis where the transformed parameters are approximately orthogonal. The two IGM parameters in this new basis are represented as ($x_0, x_1$) which are related to ($\ln \tau_0, \gamma$) as:

\begin{align}
x_0 &= -0.8563\ (\ln \tau_0 + 5.27) + 0.5165\ (\gamma - 3.21), \\
x_1 &=  0.5165\ (\ln \tau_0 + 5.27) + 0.8563\ (\gamma - 3.21).
\end{align}
The effective optical depth in terms of $x_0$ and $x_1$ is given as:
\begin{multline}
\ln \tau_\mathrm{eff} = (-0.8563\ x_0 + 0.5165\ x_1 - 5.27)\ +\\
                        (0.5165\ x_0 + 0.8563\ x_1 + 3.21) \ln (1 + z).
\end{multline}

A high degree of correlation exists between measurements of both $x_0$ and $x_1$ amongst neighboring pixels. The neighboring pixels are separated by less than one Mpc along the line of sight. Hence, we expect LSS correlations to appear in the statistics of the optical depth estimates from one restframe wavelength to another. We also attribute this correlation to the resampling of pixels and the resolution of the instrument.

To account for the effects of these processes on the statistical significance of the IGM parameter estimates, we computed the mean correlation function as a function of pixel separation for each bin. To isolate this component and ensure a positive-definite covariance matrix, each correlation function was modeled as an exponential fit to the first five data points. This exponential function is a decent fit to each mean correlation function with a typical scale length $r_0=2.24$ pixels along the $x_0$ direction and $r_0=3.4$ pixels along the $x_1$ direction. We converted the correlation function into a correlation matrix by assigning each band diagonal element the corresponding value from the analytic fit to the correlation function. The correlation matrix was transformed into a covariance matrix by weighting each element by the variance estimates defined earlier.

Incorporating the effects of covariance into the final likelihood surface is complicated by the fact that the likelihood surfaces for each restframe wavelength are not Gaussian. When ignoring the correlation between pixels and approximating the likelihood surface for each restframe wavelength as an uncorrelated two-dimensional Gaussian, we find that the confidence intervals are preserved in the combined likelihood surface when compared to a fit using the original likelihoods. However, biased estimates of the central values of $x_0$ and $x_1$ are produced. To account for the correlations and the fact that the contours for each restframe wavelength are not Gaussian, we adopted the following procedure: we first defined a stretch factor as the amount by which the uncertainty estimates change when taking the small scale correlations into account, as compared to a simple weighted estimate without correlations. This calculation was done independently along the $x_0$ and $x_1$ direction, yielding stretch factors $s_0$ and $s1$, respectively. We incorporate the effect of small scale correlations into the covariance matrix of the full likelihood estimate of $x_0$ and $x_1$ without correlations by rescaling as follows:
\[ C' = \begin{bmatrix} s_0 & 0 \\ 0 & s_1\end{bmatrix} C \begin{bmatrix} s_0 & 0 \\ 0 & s_1\end{bmatrix}.\]

The addition of cosmological correlations to the analysis has the impact of increasing the uncertainty in the $x_0$ estimates by an average factor of $s_0 = 2.04$. The uncertainty in $x_1$ estimates is increased by an average factor of $x_1 = 2.23$.  The correlation between neighboring pixels effectively reduces the number of independent restframe wavelengths by 75\%.

\begin{figure*}
\centering
\includegraphics[width=\textwidth]{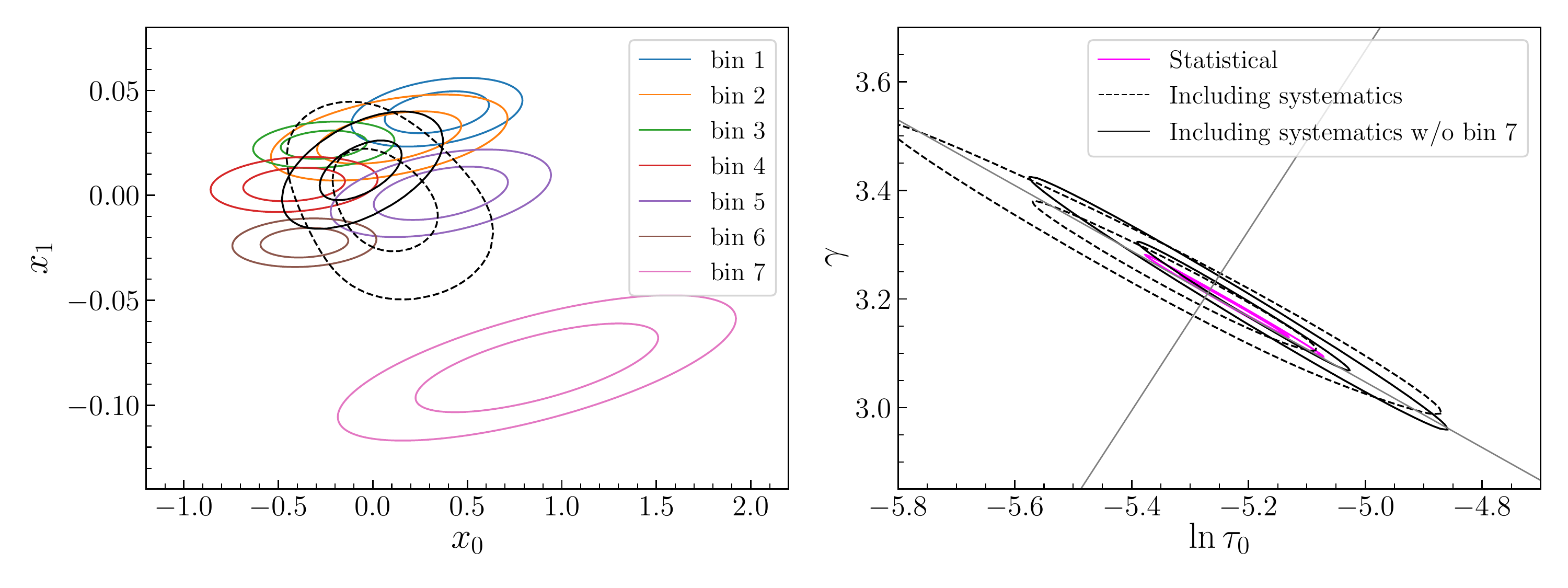}
\caption{Constraints on the optical depth parameters with $68.3\% (\Delta \chi^2 = 2.3)$ and $95.5\% (\Delta \chi^2=6.2)$ confidence regions. {\bf Left}: The black contours indicate the combined estimates, including all the bins in dashed and excluding bin 7 in solid. {\bf Right:} The combined estimate with statistical errors only is shown in magenta, while those incorporating systematic errors are represented in black with the same scheme as in the left panel. The coordinate system of the modified basis is shown with gray lines in the right panel.}
\label{fig:all_bins}
\end{figure*}

\subsection{Combined estimates} \label{sec:combined}
Figure \ref{fig:all_bins} presents the optical depth parameter estimates for all seven bins. The left panel displays the two-dimensional contours for each quasar bin in the orthogonal basis. The contours for the combined statistical likelihood from these seven measurements are shown in magenta in the right panel. The number of quasars in each bin, the best-fit estimates of $\ln \tau_0$ and $\gamma$, and their statistical errors are listed in Table~\ref{tb:best-fit}. The constraints on the parameters arising from statistical errors for each bin are similar, ranging from 2.5\% to 5.9\% on $\ln \tau_0$ and 2.5\% to 6.3\% on $\gamma$. However, the central values deviate by more than what would be expected from the reported statistical errors, as seen in the left panel of Figure~\ref{fig:all_bins}. The fifth column of Table~\ref{tb:best-fit} reports the probability that each individual measurement would produce the central value resulting from the combined likelihood. Four of the seven bins have a p-value less than one percent.

\begin{deluxetable*}{ccrrcc}
\tabletypesize{\scriptsize}
\tablecaption{Best fit values for the optical depth parameters with statistical errors\label{tb:best-fit}
}
\tablewidth{0pt}
\tablehead{
    \colhead{bin} & \colhead{$\#$ quasars} & \colhead{$\ln \tau_0$} &  \colhead{$\gamma$} &  \colhead{p-value (stat.)} & \colhead{p-value (stat. + sys.)}
    }
\startdata
bin 1 & 4625& $-5.54 \pm 0.15 $ & $3.42 \pm 0.10 $& $5.8 \times 10^{-7}$  & 0.28\\
bin 2 & 4203& $-5.33 \pm 0.21$ & $3.28 \pm 0.13$& $1.4 \times 10^{-2}$  & 0.71\\
bin 3 & 7477& $-5.04 \pm 0.13$ & $ 3.10 \pm 0.08$& $4.5 \times 10^{-6}$  & 0.69\\
bin 4 & 7715& $-4.91 \pm 0.15$ & $3.00 \pm 0.09$& $1.0 \times 10^{-1}$  & 0.51\\
bin 5 & 3855& $-5.58 \pm 0.20$ & $3.40 \pm 0.12$& $9.8 \times 10^{-2}$  & 0.73\\
bin 6 & 8824& $-4.97 \pm 0.13$ & $3.00 \pm 0.08$& $2.3 \times 10^{-8}$  & 0.36\\
bin 7 & 3336& $-6.06 \pm 0.36$ & $3.59 \pm 0.23$ & $3.1 \times 10^{-25}$ & 0.06\\
\hline
\rule{0pt}{3ex}
Coadd & 40,035 & $-5.22 \pm 0.06$ & $ 3.19 \pm 0.04$ \\
Coadd (w/o bin 7) & 36,699 & $-5.13 \pm 0.06$ & $ 3.14 \pm 0.04$ \\
\hline
\rule{0pt}{3ex}
Coadd with systematic errors & 40,035 & $-5.31 \pm 0.14$ & $ 3.23 \pm 0.09$ \\
Coadd with systematic errors (w/o bin 7) & 36,699 & $-5.20 \pm 0.11$ & $ 3.18 \pm 0.07$
\enddata
\end{deluxetable*}

We attribute the extra scatter in the optical depth parameters in each bin to systematic errors not accounted for in the analysis. We model this systematic component using a nuisance covariance matrix that is added to the statistical covariance matrix of each data point in the modified basis. The seven measurements are fit with a model that includes additional free parameters describing the systematic error on $x_0$, the systematic error on $x_1$, and their correlation.

After rotating back to the $\tau_0 - \gamma$ basis, the most likely values of the systematic covariance matrix assigned to each bin are estimated to be:
\begin{equation} \label{eq:systematics}
    C_{\mathrm{sys}}[\ln \tau_0, \gamma] =
    \begin{bmatrix}
     0.106 & -0.057 \\ -0.057 & 0.032 \\
    \end{bmatrix}.
\end{equation}

We marginalize over these nuisance parameters to obtain the maximum likelihood estimates for the optical depth parameters. The net effect of the systematic errors is to increase the uncertainty on $\ln \tau_0$ by a factor of 2.5 and on $\gamma$ by a factor of 2.3. These estimates, transformed back to the $\tau_0 - \gamma$ basis are:
\begin{align} \label{eq:best_fit}
&\tau_0 = (5.01 \pm 0.76) \times 10^{-3}, \nonumber\\
& \gamma = 3.231 \pm 0.086, \nonumber \\
&corr =0.99.
\end{align}

We then recompute the p-value for each of the seven measurements using the statistical+systematic covariance matrices. These p-values are listed in the last column of Table~\ref{tb:best-fit}. Given the sample size none of the measurements are significantly deviant.

% Approximating the full likelihood as a two dimensional Gaussian,

The 68\% and 95\% contours for the full two-dimensional likelihood are presented as dashed black curves in the left panel of Figure~\ref{fig:all_bins}. These contours overlap the 68\% statistical contours for each of bin 1 through bin 6, but do not overlap the 95\% contours in bin 7. The central value of bin 7 appears to be driving the systematic error on $x_1$ toward larger values. To determine the extent of this effect, we repeat the analysis excluding the data from bin 7. In this case, the systematic covariance matrix assigned to each bin is:

\begin{equation}
    C_{\mathrm{sys}}[\ln \tau_0, \gamma] =
    \begin{bmatrix}
     0.044 & -0.029 \\ -0.029 & 0.020 \\
    \end{bmatrix}.
\end{equation}
The variance in the systematic contribution to $\ln \tau_0$ is decreased by a factor of 2.4 after removing bin 7 and the variance in the systematic contribution to $\gamma$ is decreased by a factor of 1.6. The reduction in the size of the systematic errors alone is not sufficient evidence to remove bin~7 from the analysis; however, we provide additional evidence in Section~\ref{sec:systematics} and Section~\ref{sec:interpretation} in support of removing this bin from the analysis. Excluding bin~7, the estimates of the optical depth parameters become:
\begin{align}\label{eq:best_fit2}
&\tau_0 = (5.54 \pm 0.64) \times 10^{-3}, \nonumber\\
& \gamma = 3.182 \pm 0.074, \nonumber \\
&corr=0.98.
\end{align}
The uncertainty estimates on both $\tau_0$ and $\gamma$ are reduced by approximately 15\% after removing bin~7. The central values shift slightly towards shallower redshift evolution and higher optical depth at $z=0$. The likelihood contours after modeling the systematic errors and excluding bin 7 are shown in solid black in the $x_0 - x_1$ basis and the $\ln \tau_0 - \gamma$ basis in Figure~\ref{fig:all_bins}.

\subsection{Metal contamination} \label{sec:metals}
The mean transmission is the product of absorption from metals and Ly$\alpha$. This effect will be a function of both observer wavelength owing to the evolution of optical depth and restframe wavelength as more metal lines contribute to the absorption at shorter wavelengths. Given the low resolution of the BOSS spectra, it is impossible to correct metal absorption in each spectrum individually; instead, we assess its statistical contribution. We adopt the corrections obtained by \cite{Kirkman:2005aa} determined from 52 high resolution quasar spectra published in \cite{Sargent:1988aa}. They model the metal absorption as $DM = 1 - T = 1 - e^{-\tau_{\mathrm{eff, m}}}$:
\begin{align*}
DM(\lambda_\mathrm{rf}) = 0.01564 &- 4.646 \times 10^{-5}(\lambda_\mathrm{rf} - 1360\ \text{\AA}), \\
DM(\lambda_\mathrm{obs}) = 0.01686\ &- 1.798 \times 10^{-6}(\lambda_\mathrm{obs} - 4158\ \text{\AA}).
\end{align*}

For the evolution of metal transmission over the observed wavelength range, we evaluate the range of transmission values at a restframe wavelength $\lambda_\mathrm{rf} = 1115$ \AA. At this restframe wavelength $DM(\lambda_\mathrm{obs} = 3700\ \text{\AA}) = 0.028$ and $DM(\lambda_\mathrm{obs} = 7000\ \text{\AA}) = 0.023$, corresponding to the extremes in the redshift range of our study. The evolution only amounts to 0.5\%, sufficiently small to be neglected. This contribution to absorption from metals will lead to an underestimation of the unabsorbed continuum, $C(\lambda_\mathrm{rf})$ by $2.5\%$. When reconstructing the continuum in each bin using the optical depth parameter estimates, we correct for this contribution from metals by scaling the estimated value of the unabsorbed continuum by the inverse of the average transmission of the metals.

\section{Systematic errors}\label{sec:systematics}
We make two core assumptions in our analysis: that spectral index and equivalent width sufficiently capture the spectral diversity and that the continuum can be standardized across the entire sample of quasars in each bin. Here we investigate these assumptions.

\subsection{Different basis} \label{sec:different_basis}
Since our measurements of optical depth parameters are dominated by systematic errors, we explored other ways of binning the sample. We choose two different parameter spaces to bin the quasar population and assess the size of systematic errors compared to those found in Equation~\ref{eq:systematics}.

We first binned on spectral index and C IV FWHM because the latter was shown to be highly correlated with quasar diversity in \citet{Jensen:2016aa}. The total sample of 42,615 quasars covers seven unique bins divided by the same percentiles in spectral index and C IV FWHM as was done in Section~\ref{sec:data}. The FWHM was determined from the best-fit double Gaussian model as described in Section~\ref{sec:sample_selection}. Constructing composite spectra revealed a redshift evolution in the equivalent widths that is likely due to Malmquist bias. The equivalent widths of emission lines in the forest are small \citep{Harris:2016aa}, so this variation is not expected to be significant. We follow the procedure presented in Section~\ref{sec:framework} and Section~\ref{sec:combined} to estimate $\tau_0$ and $\gamma$ and their associated systematic errors. The best-fit optical depth parameters lie 2.8 standard deviations from those found in Equation~\ref{eq:best_fit}, indicating a shallower evolution in $\tau_\mathrm{eff}$. The systematic error covariance matrix is:

\begin{equation}
    C_{\mathrm{sys}}[\ln \tau_0, \gamma] =
    \begin{bmatrix}
     0.114 & -0.067 \\ -0.067 & 0.040 \\
    \end{bmatrix}.
\end{equation}

The variance between estimates of $\gamma$ due to systematics errors in this basis are roughly 25\% higher than in the $\alpha_\lambda -$ $W_\lambda$(C IV) basis used for the main results of this work.

We next binned on Eddington ratio and black-hole mass based on single-epoch spectroscopic scaling relationships. We used the relationship between black-hole mass, luminosity and C IV FWHM presented in \cite{Shen:2011aa}. Using this basis is an attempt to divide the sample by the physical parameters of the quasars. Although we do not bin by spectral index, we still perform a power-law correction to the continuum as in Section~\ref{sec:sample_selection}. The total sample of 42,439 quasars covers seven unique bins covering approximately the same fraction of the parameter space as in Section~\ref{sec:data}. We again follow the procedure presented in Section~\ref{sec:framework} and Section~\ref{sec:combined} to estimate $\tau_0$ and $\gamma$ and their associated systematic errors. The best-fit optical depth parameter $\tau_0$ lies 2.5 standard deviations below and $\gamma$ lies 3.9 standard deviations above from those found in Equation~\ref{eq:best_fit}. The systematic error covariance matrix is:

\begin{equation}
    C_{\mathrm{sys}}[\ln \tau_0, \gamma] =
    \begin{bmatrix}
     1.39 & -0.85 \\ -0.85 & 0.52 \\
    \end{bmatrix}.
\end{equation}

\begin{deluxetable*}{lccccccc}
\tabletypesize{\scriptsize}
\tablecaption{Percent change in continuum in the forest due to changes in spectral index on the redside\label{tb:percent}}
\tablehead{
    \colhead{Bin} & \colhead{$\bar{\alpha}$} & \colhead{$2.3< z_q \le 2.5$}& \colhead{$2.5 < z_q \le 2.7$}& \colhead{$2.7< z_q \le 2.9$}& \colhead{$2.9< z_q \le 3.1$}& \colhead{$3.1< z_q \le 3.3$} & \colhead{$3.3< z_q \le 3.5$}
}
\startdata
Bin 1 &-2.35 &-3.02 &-3.87 &-2.91 &-1.29 &-2.19 &-2.86 \\
Bin 2 &-2.37 & -1.81 &  -3.47 &-0.69 &0.36 &-0.82 &NA \\
Bin 3 &-1.75 & -1.21 &  -1.40 &1.62 &4.55 &2.52 &1.23 \\
Bin 4 &-1.78 & -0.08 &  0.22 &3.08 &4.72 &1.64 &2.69 \\
Bin 5 &-1.81 & 0.44  & 1.98 &4.14 &3.97 &3.60 &NA\\
Bin 6 &-1.21 & 0.30  & 2.06 &5.01 &6.35 &6.41 &5.13\\
Bin 7 &-1.25 & 0.00  & 3.00 &6.94 &8.58 &6.21 &NA
\enddata
\end{deluxetable*}

The variance between measurements due to systematics errors in this basis are so much larger than in the $\alpha_\lambda -$ $W_\lambda$(C IV) basis, that we infer a flaw in the assumptions of diversity based on estimates of Eddington ratio and black hole mass. Another indication for the problem in this basis is the low number of quasars above redshift $z = 3$ in several of the bins.

This exercise highlights that care is needed in identifying subsamples to have accurate optical depth measurements. The test using Eddington ratio and black hole mass to isolate diversity demonstrates that poor coverage at high redshifts leads to a model that prefers higher values of $\gamma$, at least in the case of the BOSS spectroscopic sample. Indeed, a similar trend can be seen in bins 1, 2, 5 and 7 in the main analysis. These four bins are each roughly half the sample size of the other bins. There is also a slightly lower representation at higher redshifts; 10\% of the quasar sample in bin 3 lies at $z > 3$, while only 5\% of bins 2 and 7 lies at these higher redshifts.

\begin{figure*}
\includegraphics[trim=-13 15 -50 0, width=0.5\textwidth]{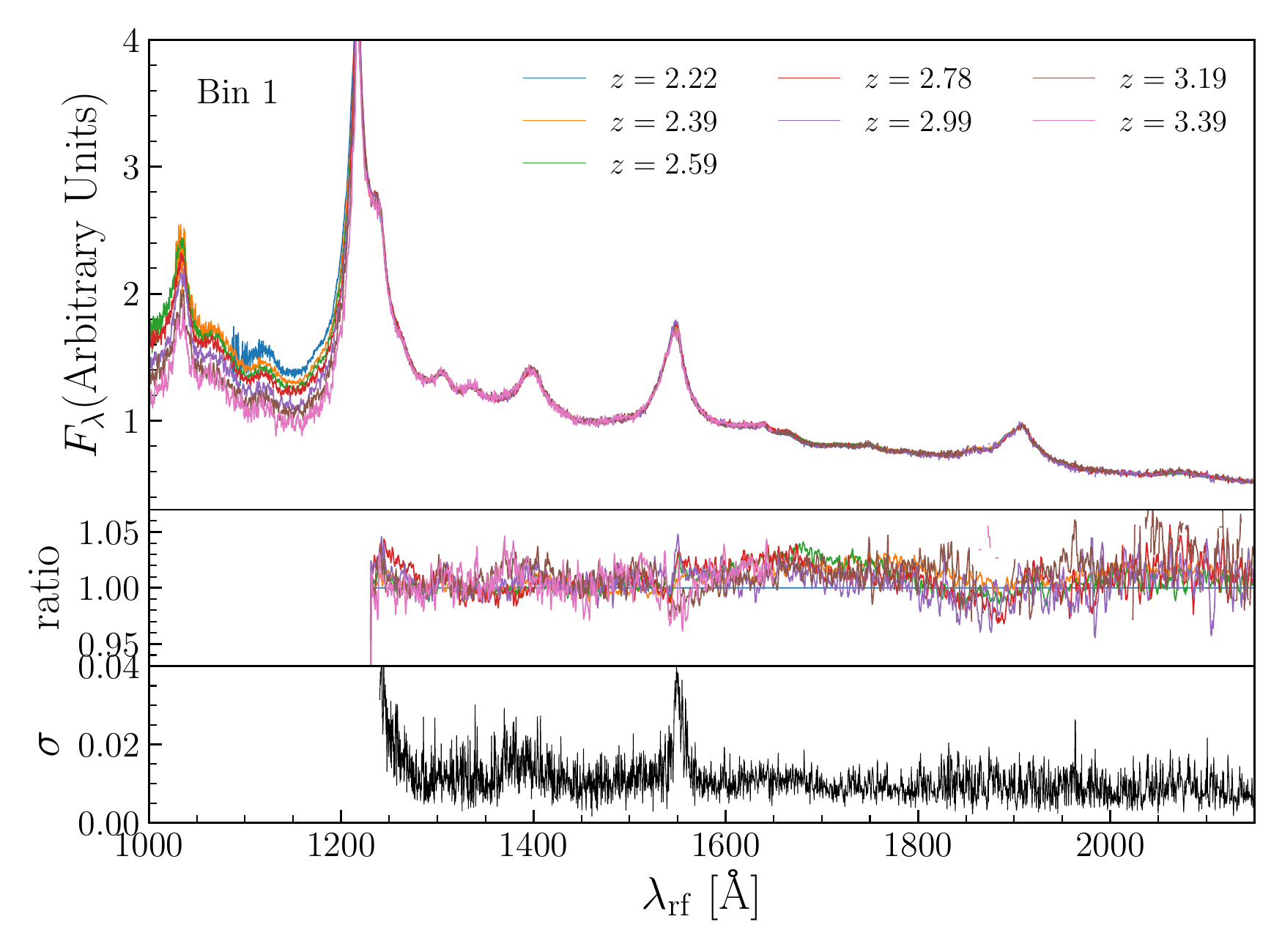}
\includegraphics[trim=-13 15 -50 0, width=0.5\textwidth]{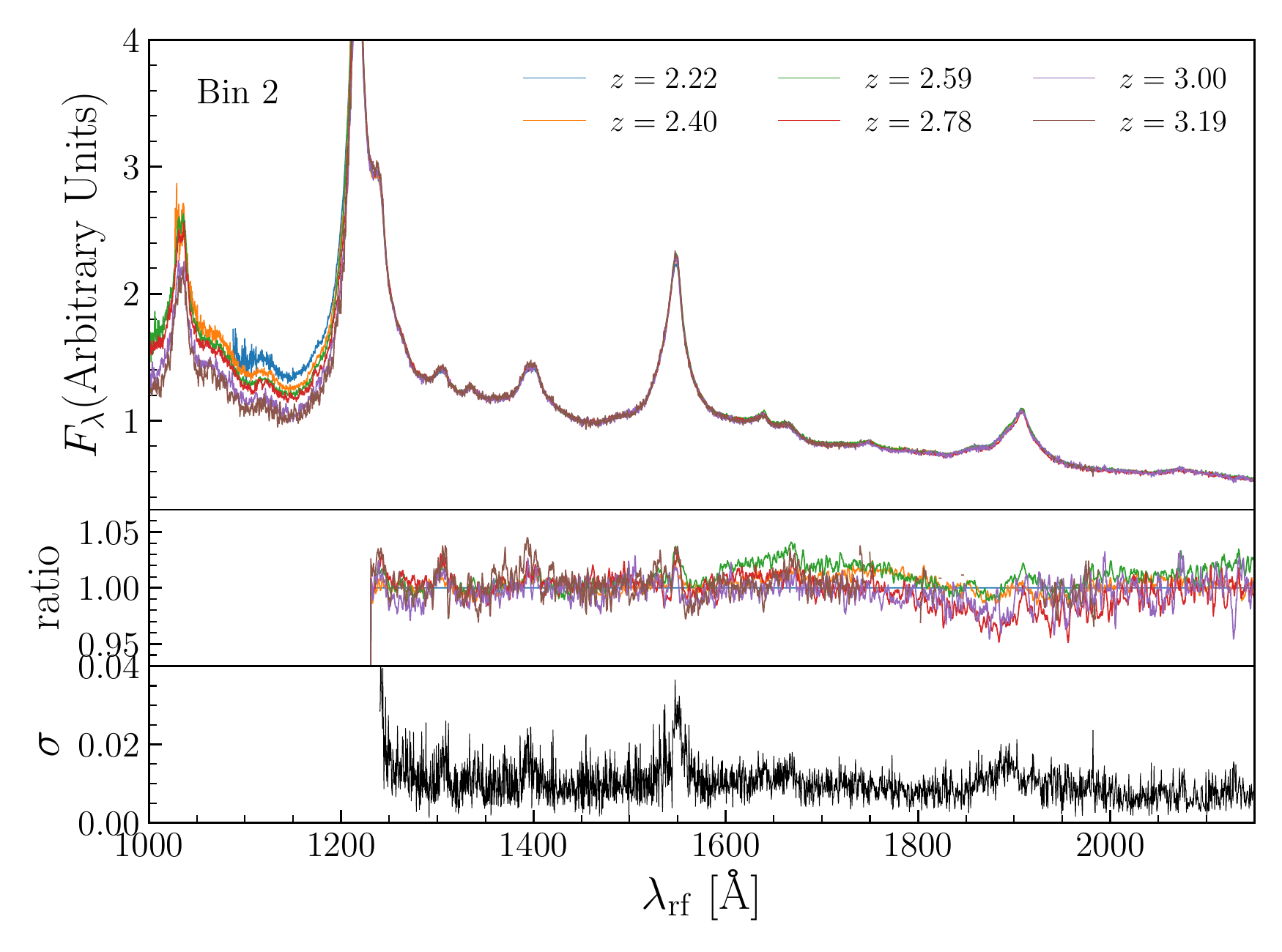}
\includegraphics[trim=-13 15 -50 0, width=0.5\textwidth]{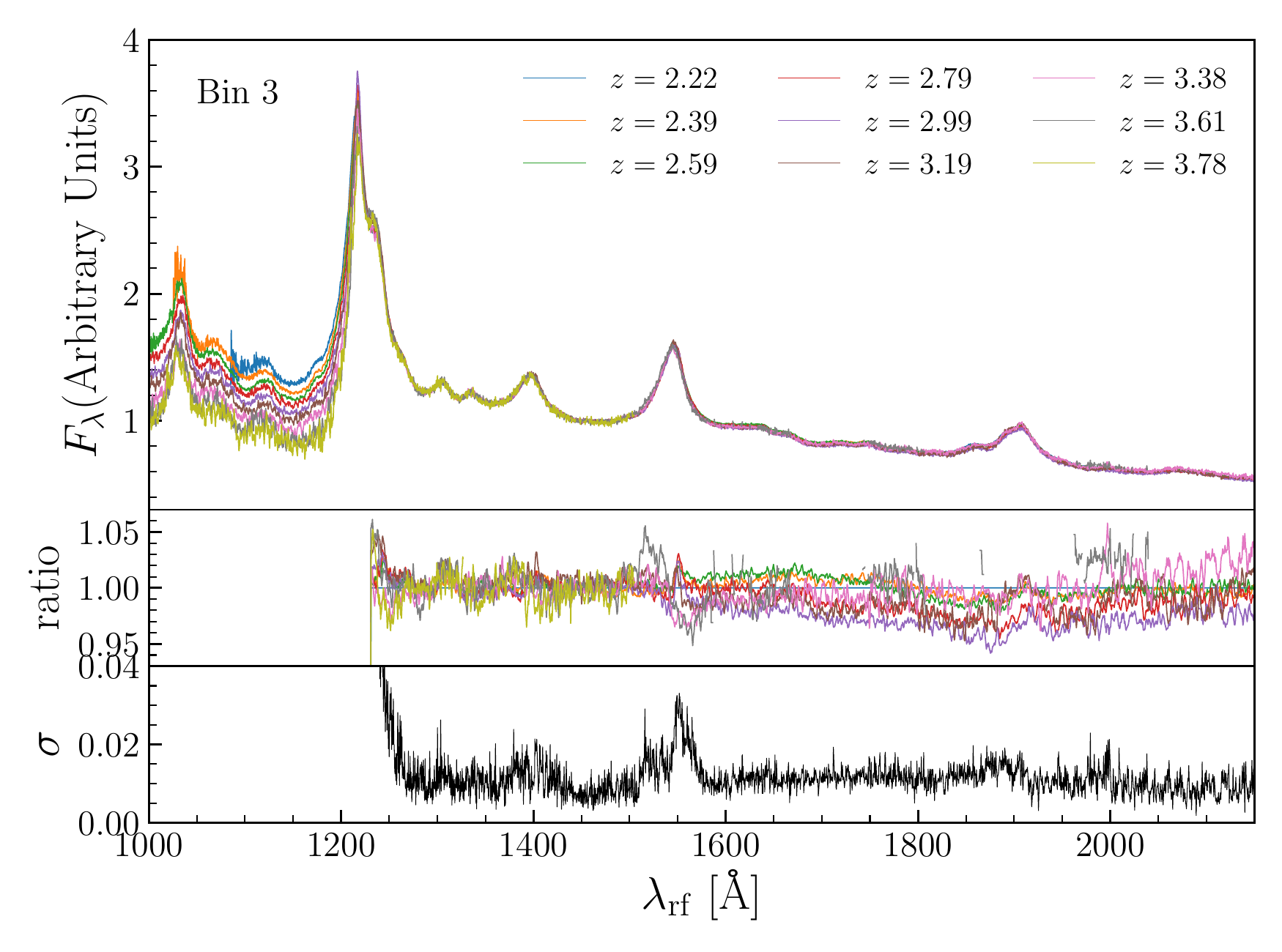}
\includegraphics[trim=-13 15 -50 0, width=0.5\textwidth]{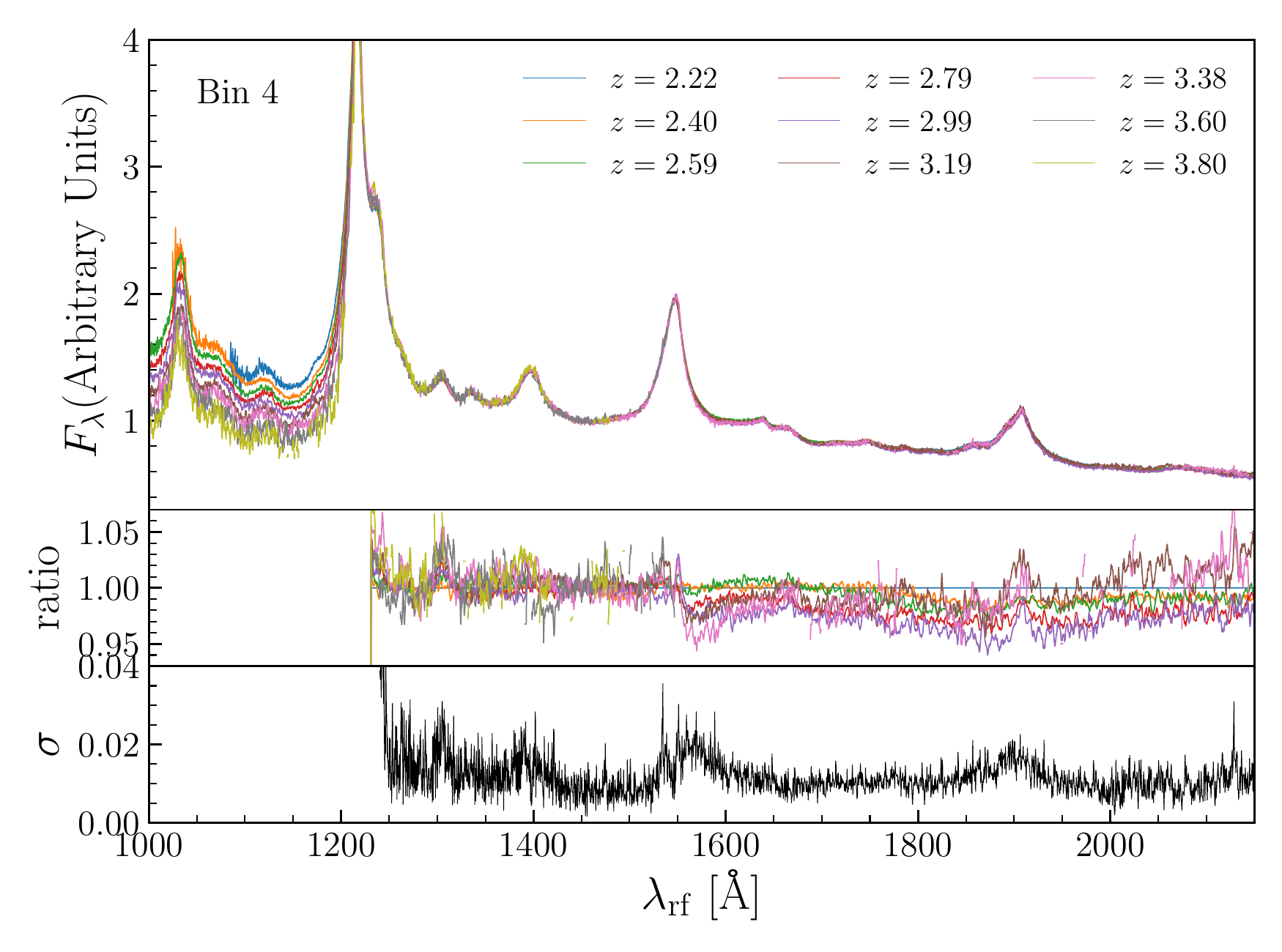}
\includegraphics[trim=-13 15 -50 0, width=0.5\textwidth]{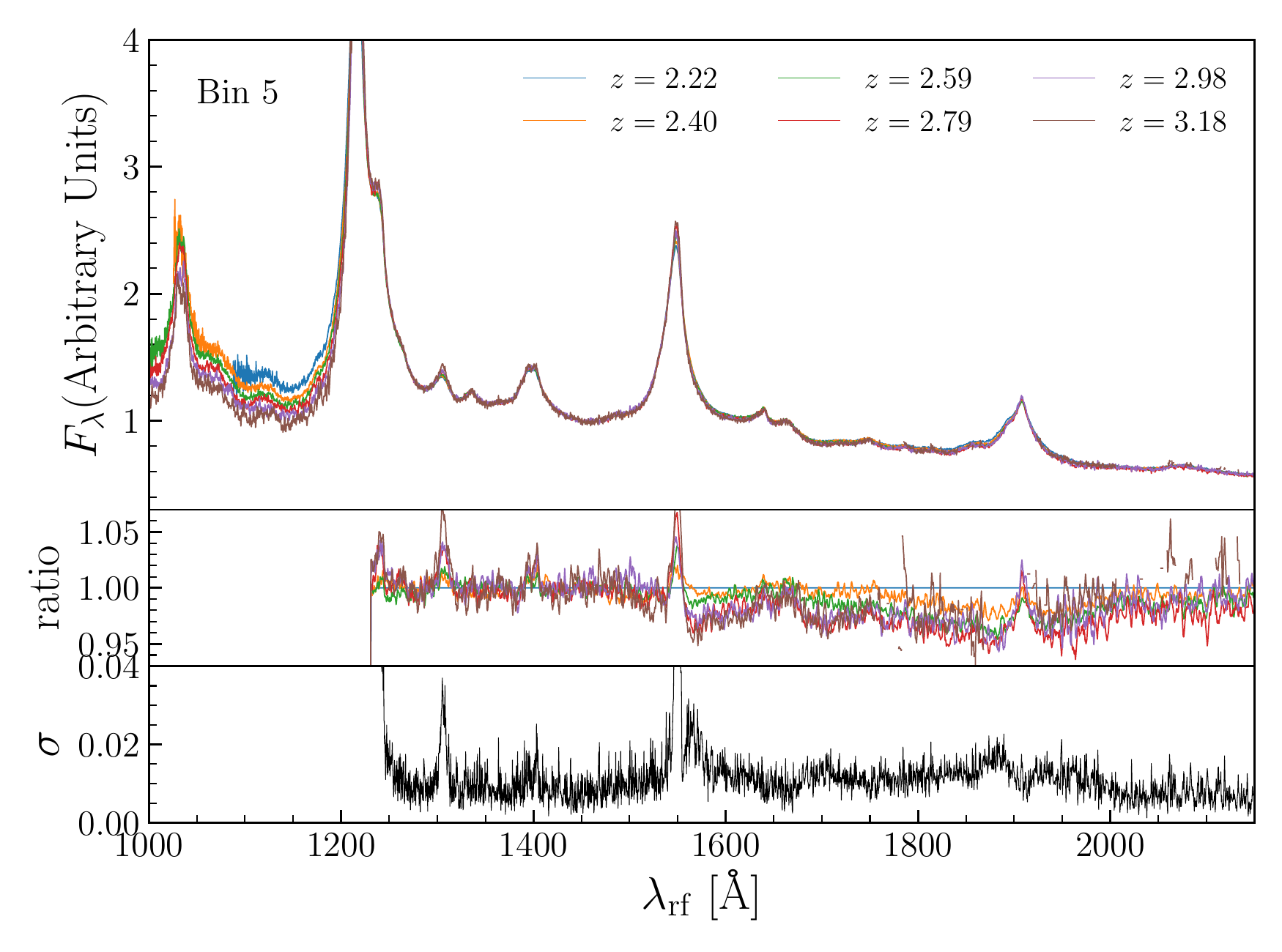}
\includegraphics[trim=-13 15 -50 0, width=0.5\textwidth]{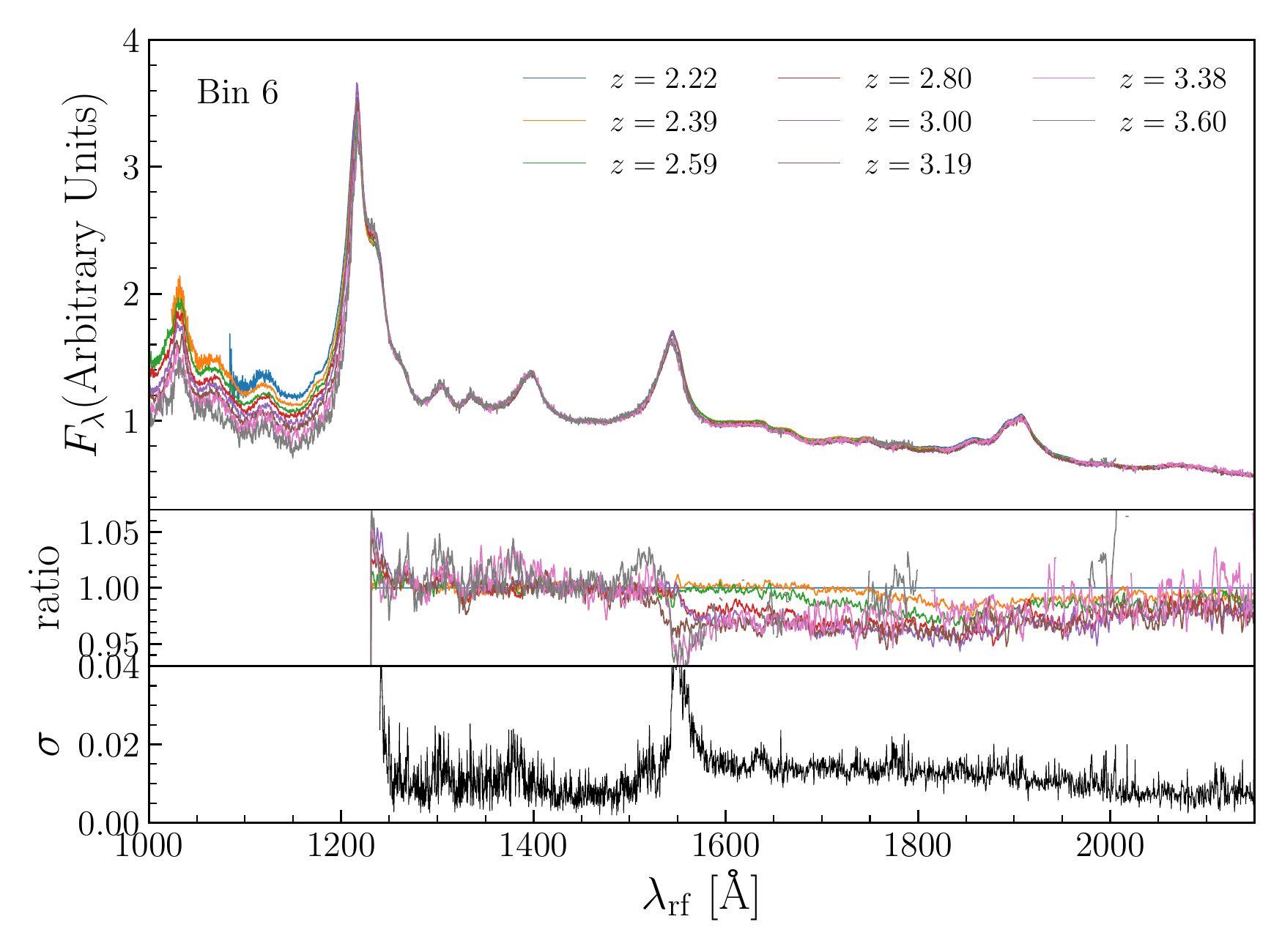}
\includegraphics[trim=-13 15 -50 0, width=0.5\textwidth]{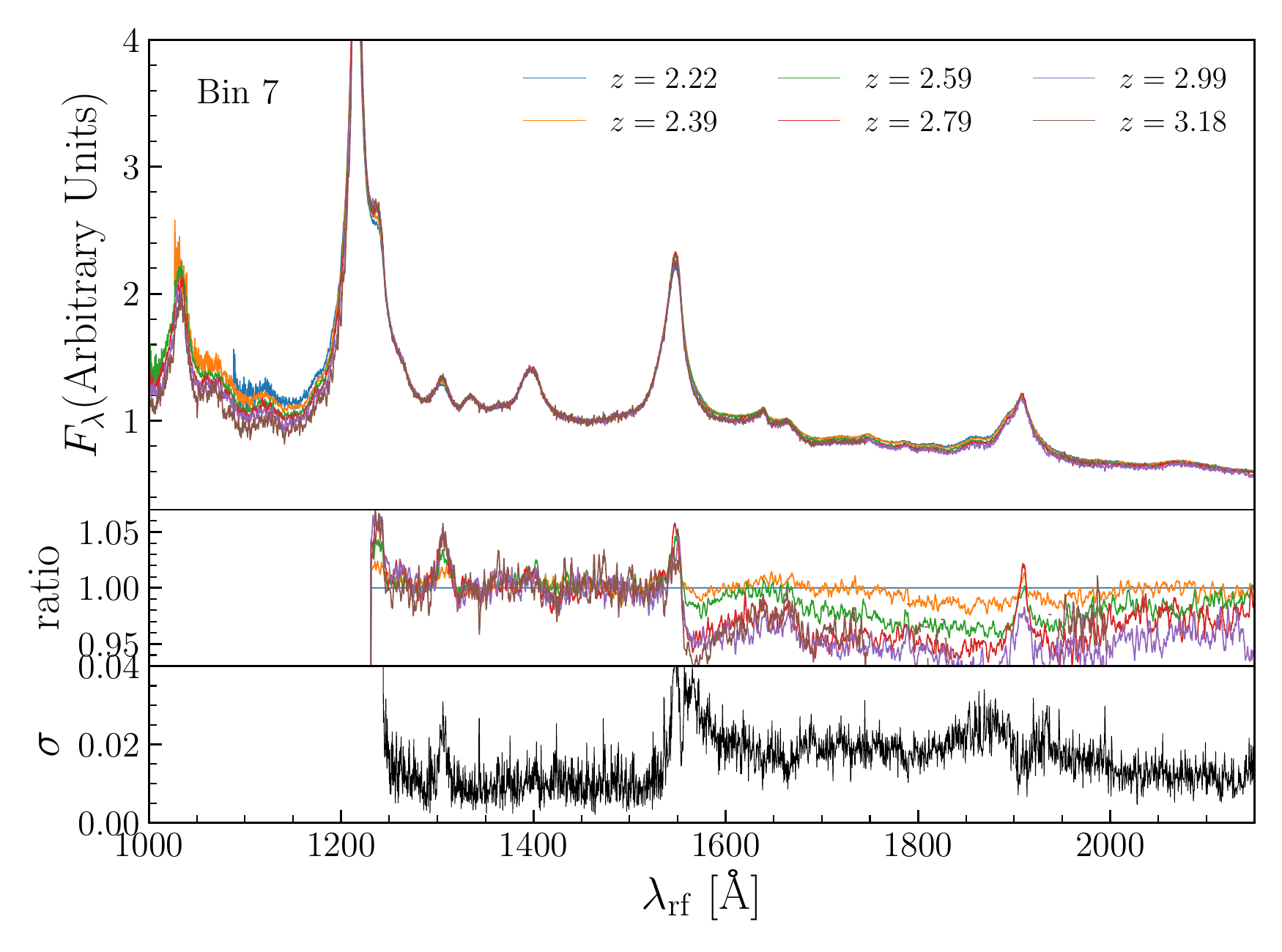}
\caption{Composite spectra for all seven bins in quasar redshift intervals of size $\Delta z_q = 0.2$. The top panel for each bin displays the normalized composite spectra, with the legend indicating the average redshift for the composite. The middle panel presents the flux ratio with respect to the composite spectrum in the first redshift interval, while the bottom panel shows the dispersion amongst the flux values redward of Ly$\alpha$.}
\label{fig:composites}
\end{figure*}

\subsection{Bias in spectral index measurements}
One possible source of error neglected in our analysis is the uncertainty in the values of spectral indices. Variations in the spectral index imply variations amongst the pixels in the Ly$\alpha$ forest that are correlated for a given quasar. Incorporating this correlated noise is complicated because data from individual quasars are modeled independently across restframe wavelengths. Here, we use a MonteCarlo approach to assess the contribution of statistical and systematic errors on spectral index to the systematic error matrix of our measurements of optical depth parameters.

The typical uncertainty in spectral index measurement for a given quasar is $\sigma_{\alpha_\lambda} \sim 0.1$. Extrapolating this uncertainty from the region used to measure the spectral index to shorter wavelengths results in an uncertainty of $\sim 3\%$ in the continuum level in the Ly$\alpha$ forest. This quantity is small compared to the dispersion introduced by the LSS term and the typical measurement uncertainty at each pixel; however, this term doesn't decrease as $\sqrt{n}$ when combining the measurements across restframe wavelengths due to its correlated nature.

To estimate the effect of correlated continuum errors across the whole sample, we use simulated power-law quasar spectra created with uncertainty in the spectral index measurements. To each spectral index, we add an extra term that is sampled from a normal distribution with zero mean and standard deviation of 0.1. We then measure the best-fit $\ln \tau_0$ and $\gamma$ values for a sample of 7000 simulated spectra. This process is repeated on 100 unique realizations. The scatter in the central values of $\ln \tau_0$ and $\gamma$ captures the covariance associated with continuum errors. The covariances account for only 6\% of the systematic covariance matrix quoted in Equation~\ref{eq:systematics}.

We now investigate the bias in spectral index measurement as a possible source of systematic error. In Section~\ref{sec:sample_selection}, we chose the relatively line-free restframe wavelength region between 1280 and 1480 \AA\ to estimate the spectral index. After warping all spectra to follow the same continuum over this wavelength range, the composite spectra divided by redshift in each quasar bin, shown in Figure~\ref{fig:composites}, reveal residual spectral diversity at restframe wavelengths longer than 1600 \AA.

To gauge the possible effect of this diversity in the Ly$\alpha$ forest continuum, we model the variations at longer wavelength according to a systematic error in the spectral index estimation. The difference in spectral index for a composite spectrum in a given redshift interval for a given bin with respect to the first redshift interval is computed as:
\begin{equation} \label{eq:delta_alpha}
\Delta \alpha_i = \log \left< \frac{f_i}{f_0} \right> / \log \left( \frac{\left<\lambda_\mathrm{rf}\right>}{1450\ \text{\AA}}\right),
\end{equation}
where $f_0$ and $f_i$ are the flux vectors for the first redshift interval and the $i^{th}$ redshift interval, respectively.
The mean flux ratio, $\left< \frac{f_i}{f_0} \right>$, is computed using these flux vectors over the restframe wavelength range $1600-1800$~\AA\ and $\left<\lambda_\mathrm{rf}\right> = 1700\ \text{\AA}$. The relative change in the Ly$\alpha$ forest continuum compared to the composite spectrum in the first bin is approximated by evaluating the quantity in Equation~\ref{eq:delta_alpha} at restframe wavelength of 1100 \AA.

Table~\ref{tb:percent} presents the percent change in the Ly$\alpha$ forest continuum values if these redshift-dependent variations at 1700~\AA\ were present at 1100~\AA. To predict the systematic errors on $\tau_0$ and $\gamma$, we simulate composite spectra following the inverse variance and composite redshifts of bin 3, using our central values of $\ln \tau_0$ and $\gamma$. The continuum was perturbed at each redshift interval following the measured changes quoted in the table for bin 3. The mean optical depth parameters across all the restframe wavelengths were then measured on these simulated composite spectra. The deviations of the measured values from the input values are much larger than the systematic errors given in Equation~\ref{eq:systematics}. These results indicate that the variations observed at 1700~\AA\ cannot be used to predict the absolute variations in the Ly$\alpha$ forest continuum. However, this result does not imply the extrapolated continuum is free of redshift dependent biases.

To quantify the sensitivity of the optical depth model to redshift dependent errors in the spectral index measurements, we investigate the effect under a set of simple assumptions. We assume that the continuum for the first redshift bin ($z_q \sim 2.1$) is correctly modeled, and assume the systematic offset in spectral index measurements to be linearly increasing with redshift. This variation is parameterized by a slope parameter, $m$, that represents the fractional change in spectral index per unit redshift:

\[m = \frac{d \log \alpha}{dz}.\]

As before, we measure the mean optical depth parameters across all the restframe wavelengths on these simulated composite spectra. The offsets in $\tau_0$ and $\gamma$ are compared to the input values as a monotonic function of $m$. As the slope increases, the central value of $\tau_0$ decreases and vice versa for $\gamma$. We focus on bin 7 because it is the largest outlier in optical depth parameters in Table~\ref{tb:best-fit} and in continuum offsets in Table~\ref{tb:percent}. The central value of $\ln \tau_0$ measured in bin 7 differs from the global central value by 0.75 and the central value of $\gamma$ measured in bin 7 differs from the global central value by 0.36. A fractional change in the spectral index per unit redshift as large as 7.6\% is required to explain this shift in the optical depth parameters.

\section{Interpretation} \label{sec:interpretation}
In this section, we consider the measurements of the optical depth parameters presented in Section~\ref{sec:optdepth} in the context of predictions for the unabsorbed continuum. We examine the diversity of features in the Ly$\alpha$ forest of the quasar spectra. We then present the measurements of the mean transmission from the observer frame to test how accurately we predict the underlying continuum.

\subsection{Reconstructed continuum} \label{sec:reconstruction}
\begin{figure}
\centering
\includegraphics[width=0.45\textwidth]{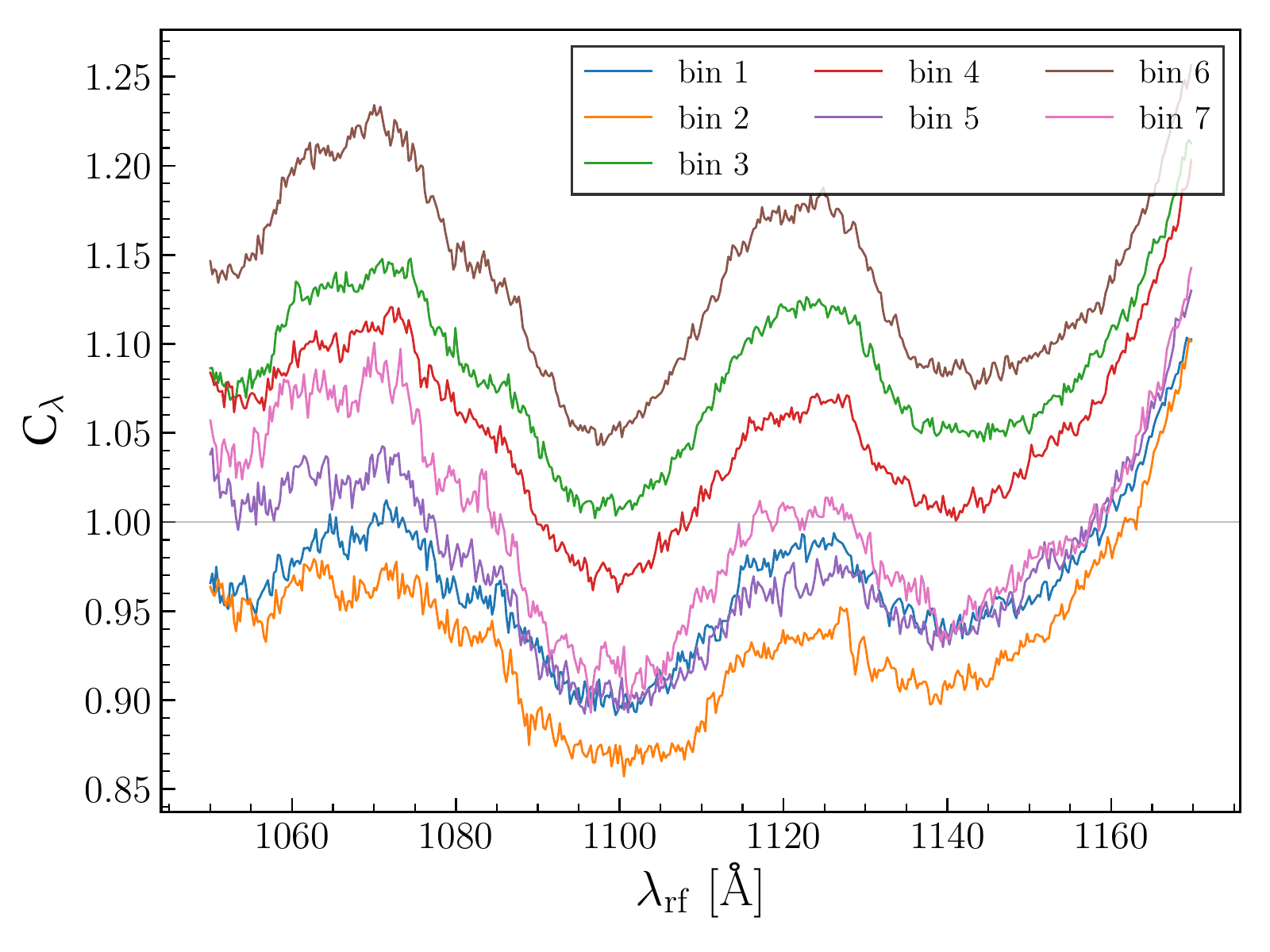}
\includegraphics[width=0.45\textwidth]{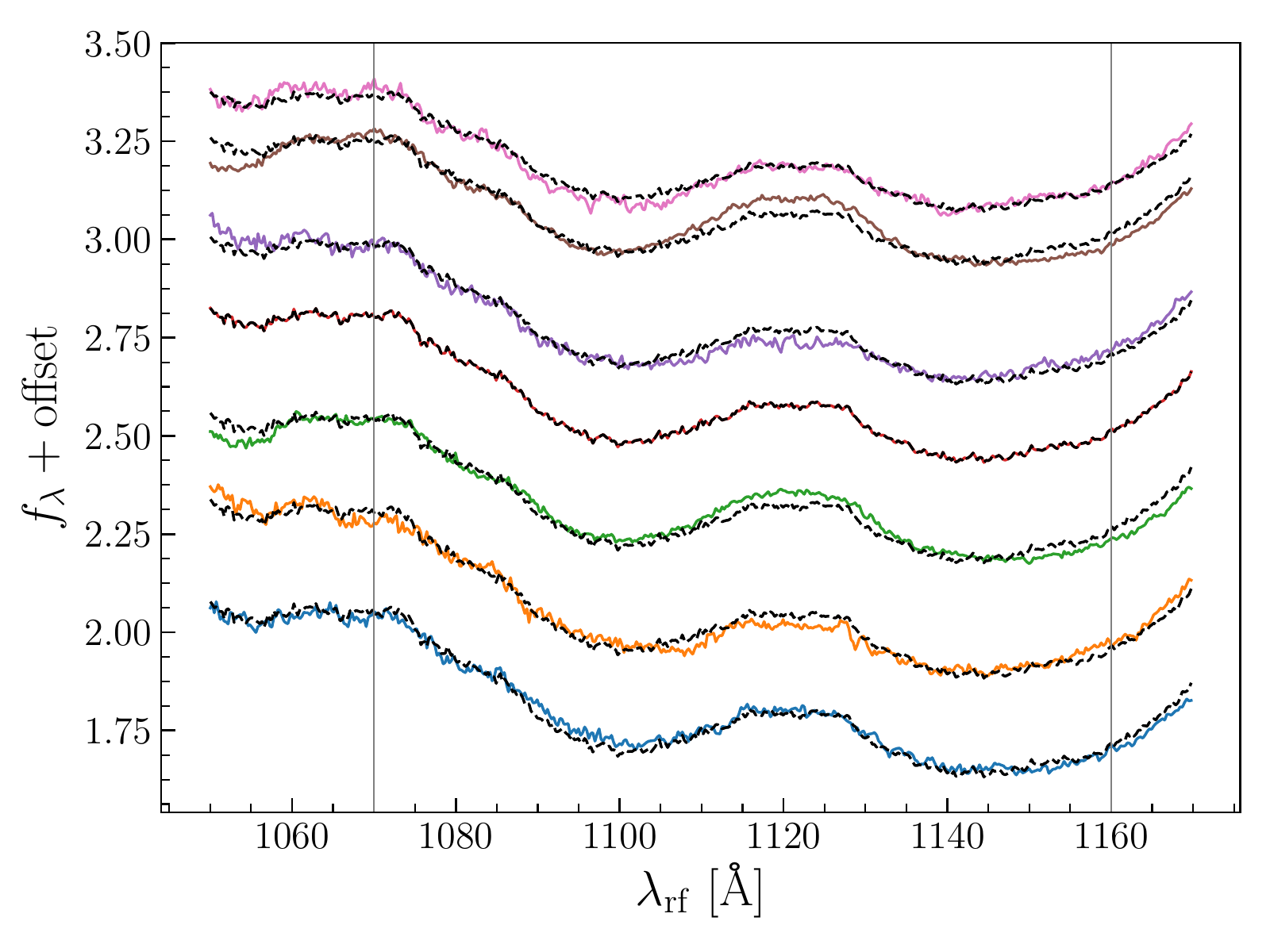}
\caption{{\bf Top:} Reconstructed quasar continuum in the Ly$\alpha$ forest for each bin using the respective best-fit optical depth parameters. The power-law continuum using the mean value of the spectral index for each bin has been extrapolated to the Ly$\alpha$ forest region and removed. {\bf Bottom:} The continuum of each of the seven bins compared to a warped continuum from bin 4 shown in black. The color scheme is the same as the top panel and the curves are offset in increments of 0.3 along the flux density axis for illustrative purposes.}
\label{fig:reconstructed}
\end{figure}

% We estimate the unabsorbed continuum in the Ly$\alpha$ forest region. From this model, we assess how well we predict the absolute continuum level and assess diversity in the line features across the whole sample of quasars.
We reconstruct the Ly$\alpha$ forest continuum for each quasar spectrum using the best-fit optical depth parameters from its respective bin in Table~\ref{tb:best-fit}; then apply the same spectral warping to each spectrum as was done in Section~\ref{sec:framework}. A composite spectrum for each bin is created from these unabsorbed spectra. We correct for the contribution of metals as discussed in Section~\ref{sec:metals}. The resulting composite spectra covering the forest range $1050 - 1170$ \AA\ are shown in Figure~\ref{fig:reconstructed}.

The top panel shows the reconstructed continuum normalized by a power-law extrapolation using the mean spectral index for each bin. If the estimates of spectral index were unbiased and the same power-law describes the continuum in the Ly$\alpha$ forest, then the continuum level for all composite spectra would be unity. Instead, using 1100 \AA\ as a proxy for the line-free continuum, the reconstructed spectra deviate over a range of -10\% to +5\%. The deviations from bins 1, 2, 5 and 7 indicate an underestimate of the continuum level. As stated in Section~\ref{sec:different_basis}, these bins have less representation at higher redshifts and are best described with a steeper evolution of optical depth. If the suppression of estimated continuum level is consistent across all redshifts, then no systematic bias in optical depth parameters should arise in these bins. Conversely, it is possible that the variations in predicted continuum levels are an artifact of redshift or signal-to-noise ratio dependent errors in the spectral index. Further investigation with larger, deeper samples is required to disentangle these possible causes.

The bottom panel of Figure~\ref{fig:reconstructed} presents spectral diversity amongst the bins using bin 4 as a reference. Doing so highlights the diversity in line features. The reconstructed composite spectrum from bin 4 is warped using a powerlaw and fit to each of the other six bins. The reconstructed continuum in each bin is shown in color following the same scheme as the top panel, relative to the model based on bin 4 in black. The warped model matches remarkably well with the reconstructed continuum for the other bins. The maximum difference occurs near the Fe III blend centered around 1123 \AA. For bins 3 and 6, the differences are only 2.2\% and 2.7\%, respectively, indicating a mild negative correlation with the strength of the C IV line. These results indicate that the assumptions of a uniform model for the features in the Ly$\alpha$ forest continuum made in BOSS cosmology studies \citep{Bautista:2017aa,du-Mas-des-Bourboux:2017aa} are reasonable.

\begin{figure*}
\centering
\includegraphics[width=0.95\textwidth]{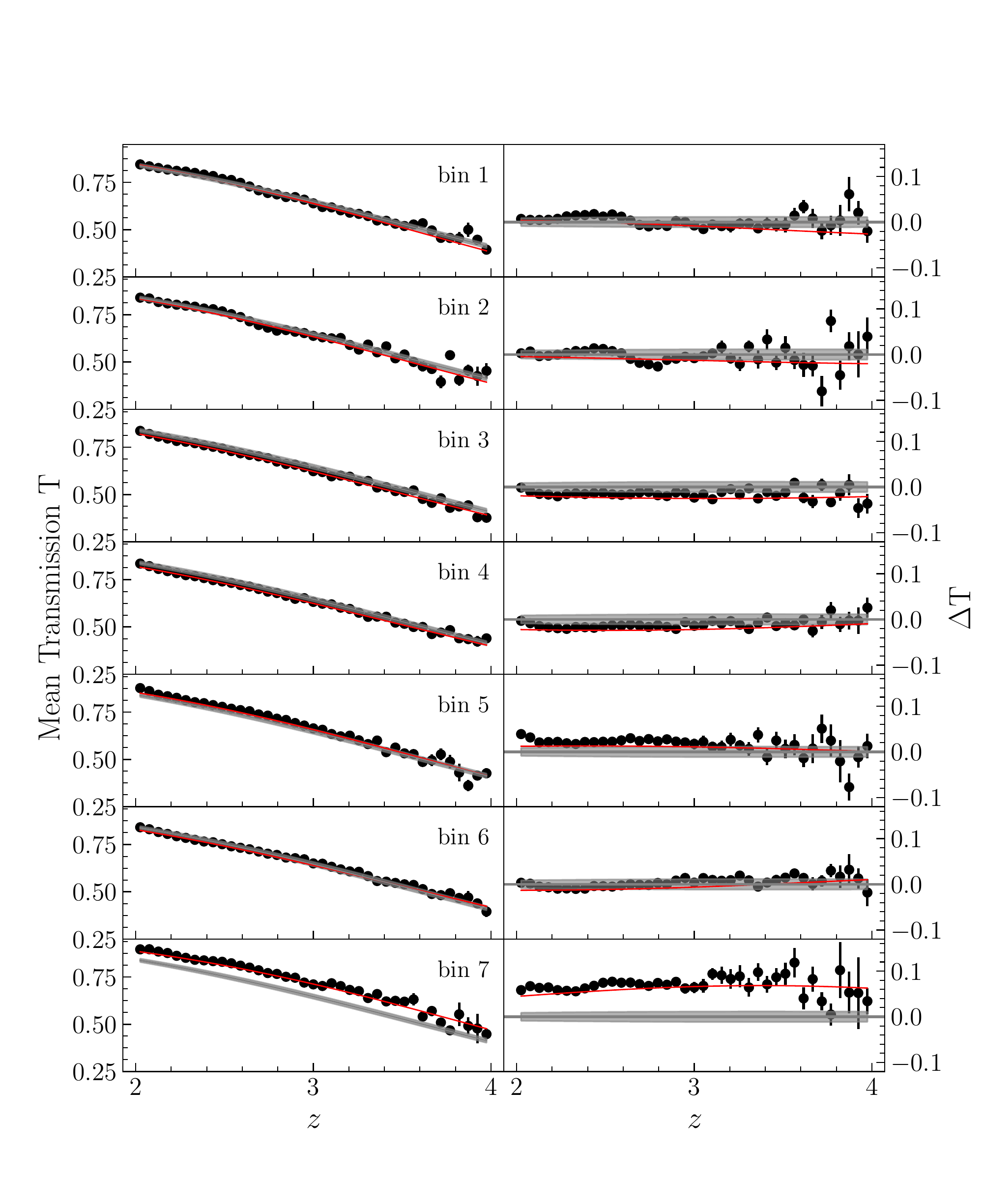}
\caption{{\bf Left:} Estimates of the transmission using the reconstructed continuum estimates for each bin as shown in Figure~\ref{fig:reconstructed}. {\bf Right:} The residuals of the measured transmission relative to the global best-fit model. In both columns, the data are shown as filled black circles with respective errorbars. The best-fit power-law for each bin is indicated in red. The overall best-fit model given by Equation~\ref{eq:best_fit} is shown in gray area, with the band giving the corresponding 1$\sigma$ uncertainty estimate.}
\label{fig:data_points}
\end{figure*}

\subsection{Mean Transmission} \label{sec:transmision}
We now revisit the analysis from the perspective of the observer frame. Our goal is to assess how well the direct estimates of mean transmission follow the power-law model for the evolution of optical depth and how the data from each bin compare to predictions from the global model.

The model in Section \ref{sec:framework} and \ref{sec:combined} leads to degeneracy between the optical depth parameters $\tau_0$ and $\gamma$, and the estimate of the unabsorbed continuum at each restframe wavelength. The work presented in Section~\ref{sec:reconstruction}, however, uses the joint estimates of $\tau_0$ and $\gamma$ from all the restframe wavelengths to provide a high-precision estimate of the reconstructed continuum for each bin. Using this continuum as a model, we measure the transmission directly as a function of redshift for each quasar in the Ly$\alpha$ forest region over the observed wavelengths. The left panels of Figure~\ref{fig:data_points} show the mean of these transmission values as a function of redshift, in bins of size $\Delta z = 0.05$.

The models using the best-fit parameters for each bin are indicated in red in Figure~\ref{fig:data_points}. Using 10,000 bootstrap resamplings to assess the covariance between data-points, we compare each model to its respective data sample. We find a typical $\chi^2 \sim 140$ for a total of 40 data-points, indicating a discrepancy between the data and the power-law model. The break from a power-law is evident in bins 1 and 2 in the form of a wiggle with a maximum excursion around $z=2.5$. However, the signature of the excursion changes across all the bins. The variable nature of the excursions indicate that the high $\chi^2$ values are not likely due to flux calibration errors or a consistent failure in the power-law model to describe the data. We were unable to identify the true source of these deviations but hypothesize that residual quasar diversity across the $2 < z < 4$ redshift range is responsible. For example, the inconsistencies at restframe wavelength 1700 \AA\ across quasar redshifts, as seen in Figure~\ref{fig:composites}, could produce such features in the measured transmission if indeed they are present at wavelengths shorter than 1200 \AA.

The right panels in Figure~\ref{fig:data_points} present the difference between the measured transmission and the global best fit model as given by the parameters in Equation~\ref{eq:best_fit}. For bins 1 through 6, the average deviations range from 0.2\% to 2.4\%, compared to a typical uncertainty of 1.8\% on the modeled transmission. Bin 7 deviates by an average of 11.5\%; indicating that the continuum is systematically underestimated when using the optical depth parameters derived from that sample of quasars ($\ln \tau_0=-6.06, \gamma=3.59$). Bin 7 was also the largest outlier in optical depth parameter estimates (Figure~\ref{fig:all_bins}) and showed the largest trend in the continuum residuals at restframe wavelengths around 1700 \AA\ (Figure~\ref{fig:composites}). This overall outlier behavior of the quasar spectra in Bin 7 is likely a result of systematic errors in the estimates of the spectral index or uncontrolled diversity in the sample. Our final constraints on the optical depth parameters henceforth (and in the abstract) are reported using only bins 1 through 6.

\section{Comparison to Previous results} \label{other_results.pdf}
This section compares our results to those from previous studies. We include works from both high-resolution and low-resolution spectroscopy in our comparison. We first compare the smooth evolution of optical depth parameterized by a power-law, then discuss whether the BOSS data provide evidence for the previously-reported He~II feature.

\subsection{Optical depth evolution}
\begin{figure*}
\centering
\includegraphics[width=0.48\textwidth]{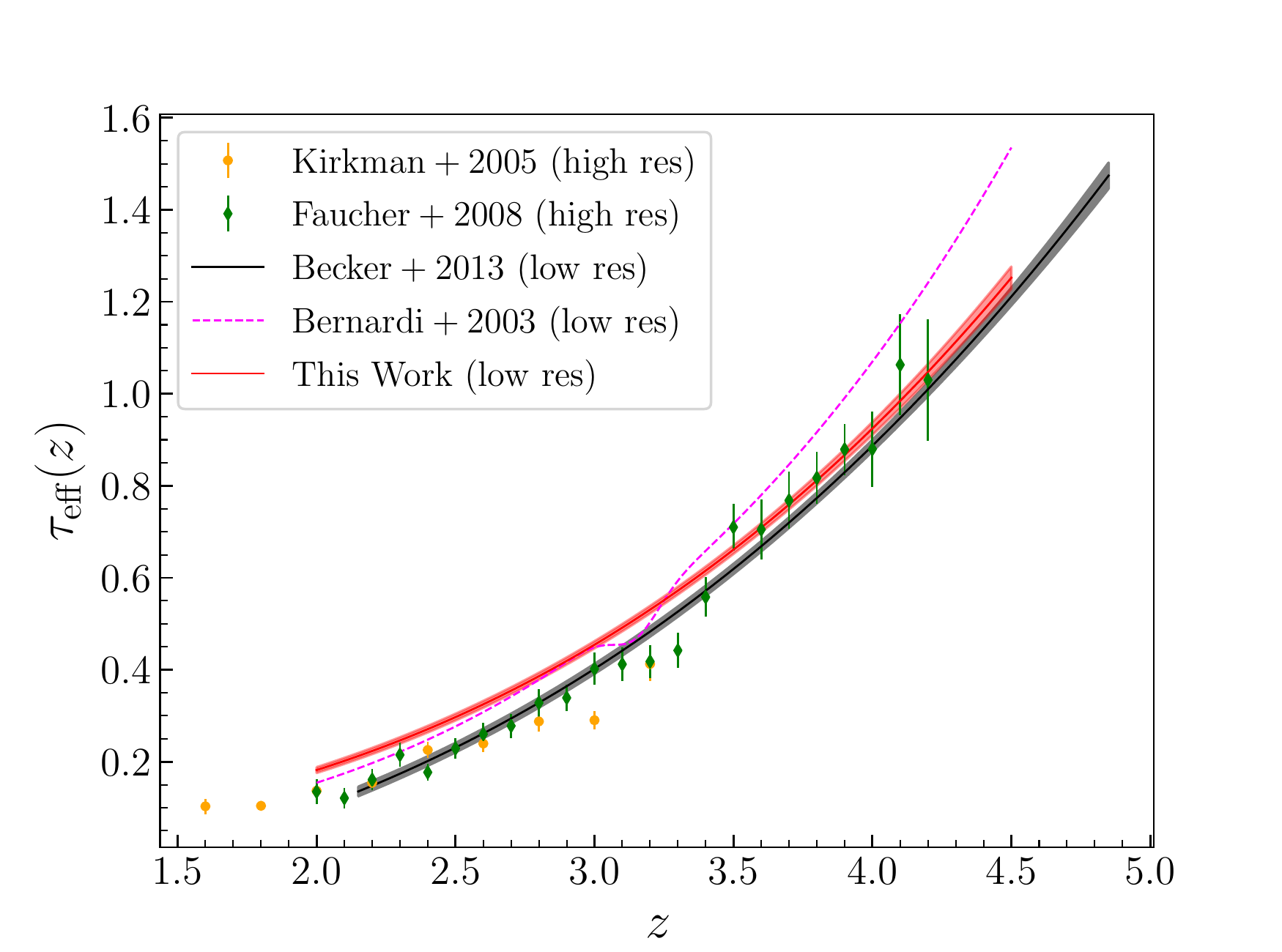}
\includegraphics[width=0.48\textwidth]{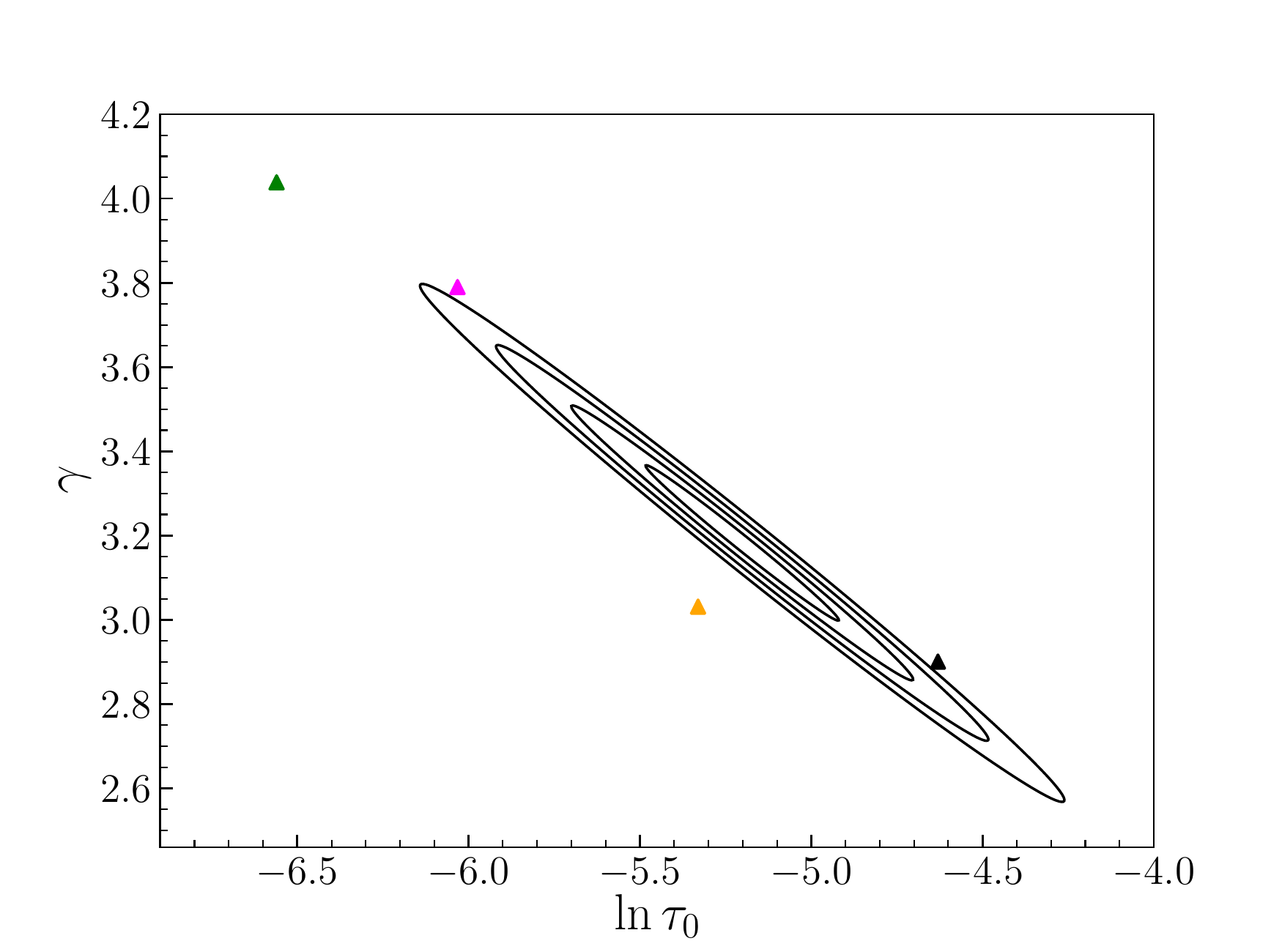}

\caption{Comparison of our measurements with published results. {\bf Left:} The evolution of the effective optical depth using the global best-fit parameters given by Equation~\ref{eq:best_fit2} is shown in red. The shaded region, also in red, is the 1$\sigma$ uncertainty in our best-fit model, after taking the covariance between the model parameters into account. {\bf Right:} The 2, 4, 6 and 8$\sigma$ confidence intervals for our measurements in the $\tau_0-\gamma$ plane are displayed in black. A Gaussian form of the likelihood to these high $\sigma$ confidence levels has been assumed. The central values of the other studies are represented in solid triangles with the same color scheme as in the left panel.}
\label{fig:other_results}
\end{figure*}

Measurements of the effective Ly$\alpha$ forest optical depth in the literature fall into two camps. High-resolution, high signal-to-noise ratio spectra allow for a direct but approximate continuum fitting using the peaks in the forest region. This method suffers from a potential underestimation of the continuum, especially at higher redshifts. However, the approach allows one to directly measure the transmission probability distribution function and the associated statistics of the transmission field. Studies such as this work that rely on low-resolution, low $S/N$ spectra typically contain more objects and therefore a potential for higher precision statistical estimates. We present a summary of our optical depth measurements relative to four other works in Figure~\ref{fig:other_results} and Table~\ref{tb:other_results}.

\cite{Kirkman:2005aa} obtained a sample of 24 high resolution spectra from HIRES spectrograph \citep{Vogt:1994aa} on the Keck Telescope. They measured the evolution of transmission over the redshift range $2.2 < z < 3.2$, after removing contributions from Lyman Limit systems (LLS) and metal lines, and attempted to remove biases due to continuum fitting by using artificial quasar spectra. Their model differs from that used in this work, in that they assume a power-law in the quantity $DA = 1 - e^{-\tau_\mathrm{eff}}$. Table~\ref{tb:other_results} presents constraints on the optical depth parameters produced by the model used in this paper, assuming their measurements shown in Figure~\ref{fig:other_results} to be independent.

A detailed analysis using a larger sample of high-resolution quasar spectra was performed by \cite{Faucher}. They measured the effective optical depth over the redshift range $2 \le z \le 4$ from a sample of 86 high-resolution, high-$S/N$ quasar spectra obtained with the ESI \citep{Sheinis:2002aa} and HIRES spectrographs on Keck and with the MIKE spectrograph \citep{Bernstein:2003aa} on Magellan. Their study contained a detailed analysis of systematic errors from continuum fitting and metal corrections. Fitting the data with a two parameter power-law produced a $\chi^2 = 40.1$ for 21 degrees-of-freedom. Adding more free parameters in the form of a three-component Gaussian, they improve the fit to a $\chi^2 = 26.8$, providing positive evidence for the additional free parameters according to the Bayesian Inference Criterion (BIC). The evidence for an expanded functional form could be attributed to He II reionization, a general deviation from a power-law, or systematic errors in the approach.

\cite{Bernardi03} employed 1061 low-resolution SDSS quasar spectra to map the redshift evolution of optical depth over the Ly$\alpha$ redshift range $2.5 \le z \le 4$. They constructed composite spectra in bins of width $\Delta z_q = 0.2$ after normalizing each spectrum by its flux in the restframe wavelength range $1450-1470$ \AA. They adopted a parametric form to describe the continuum in the forest consisting of a power-law and Gaussian features to account for emission lines at 1073 \AA, 1123 \AA\ and Ly$\alpha$ 1215.67 \AA. In addition to a smooth power-law evolution, they reported a clear `bump' at $z \sim 3.2$ that they attribute to He II reionization. Their best-fit model, including the modeled reionization feature, is shown as the brown dashed line in Figure~\ref{fig:other_results}.

A study that closely resembles our work is that of \cite{Becker:2013aa}, who used 6065 low resolution quasar spectra over the quasar redshift range $2 \le z \le 5$ from the SDSS DR7 quasar catalog \citep{Schneider:2010aa}. They constructed composite spectra in bins of quasar redshifts with a typical $\Delta z_q = 0.1$, after correcting for differences in the spectral indices. Using the normalized transmitted flux measurements in bins of $\Delta z = 0.1$, they fit the optical depth parameters simultaneously across all restframe wavelengths. To obtain absolute measurements of transmission, they scale their results to those from \cite{Faucher} at $z \sim 2.35$ using an additional free parameter. This approach leads to an additional offset parameter in the effective optical depth parameterization given as:
\[\tau_\mathrm{eff}(z) = \tau_0 \left(\frac{1+z}{1 + z_0} \right)^\gamma + C\]
They estimate the value of this offset to be $ C = -0.132$. For comparison, our most discrepant measurement in bin 7 would require a value of $C = 0.097$ to bring the measured optical depth into alignment with the global model. Their best fit three-component model for the effective optical depth is indicated in black with 1$\sigma$ confidence intervals in Figure~\ref{fig:other_results}. They reported no evidence for a He II reionization feature.

\begin{deluxetable}{lrcc}
\tabletypesize{\scriptsize}
\tablecaption{Previous measurements of the effective optical depth parameters \label{tb:other_results}}
\tablehead{
    \colhead{Work} & \colhead{\# quasars} &\colhead{$\tau_0$} & \colhead{$\gamma$}
}
\startdata
Kirkman+2005\footnotemark[1] & 24  & $0.0049 \pm 0.0011$ & $3.03 \pm 0.17$ \\
% Kim+2007 & 18 &  $0.0054 \pm 0.0101$ & $2.96 \pm 0.83$ \\
Faucher+2008 & 86 & 0.0018 & 3.92 \\
Bernardi+2003 & 1061 & $0.0024 \pm 0.0014$ & $3.79 \pm 0.18$ \\
Becker+2013 & 6065 & $0.0097 \pm 0.0021$ & $2.90 \pm 0.12$ \\
This Work & 40,035 & $0.0055 \pm 0.0006$ & $3.18 \pm 0.07$
\enddata
\footnotetext[1]{Calculated using their errorbars on data, but using our model}
\end{deluxetable}

Our best-fit model for the effective optical depth including the $\sigma$ confidence interval is shown in red in Figure~\ref{fig:other_results}. Our model predicts a larger typical opacity due to Ly$\alpha$ at $z < 3.3$ than the measurements of all the other works. Over this redshift range, the average discrepancy ranges from $\Delta \tau_\mathrm{eff} = 0.01$ in case of \cite{Bernardi03} to $\Delta \tau_\mathrm{eff} = 0.09$ in the case of \cite{Kirkman:2005aa}. We predict lower opacity than \cite{Bernardi03} by an average of $\Delta \tau_\mathrm{eff} = 0.05$ over the redshift range $3.3 < z < 4.2$. At these higher redshifts, we agree with the measurements of \cite{Faucher} and \cite{Becker:2013aa} at a level better than $\Delta \tau_\mathrm{eff} = 0.04$.

A more quantitative comparison of the best-fit models is presented in Table~\ref{tb:other_results}. At face value, our constraints are similar to those of \cite{Kirkman:2005aa}, in that the $\tau_0$ and $\gamma$ estimates are consistent to better than one standard deviation. Likewise, our measurements lie within two standard deviations of the \cite{Becker:2013aa} $\tau_0$ and $\gamma$ estimates. However, simply comparing the marginalized estimates of each parameter independently neglects the high degree of correlation between the parameters, as shown in the right panel of Figure~\ref{fig:other_results}.
% In the last column of Table~\ref{tb:other_results}, we report the probability of finding the central values of each study given our $\tau_0 - \gamma$ likelihood surface.
% To compute this, we add the likelihoods in quadrature assuming the same correlation found in our study.
There is a large disagreement between all of the studies and our own. The measurement of \cite{Becker:2013aa} and \cite{Bernardi03} are in closest agreement, although they lie right at the edge of our 8$\sigma$ contour. The prediction of steeper evolution compared to our work as found in \cite{Faucher} and \cite{Bernardi03} indicates that their estimates lie in the extremes of our $\tau_0 - \gamma$ degeneracy curve.

\subsection{He II feature}
A few previous studies \citep{Bernardi03, Faucher} report detection of a narrow feature superposed on a smooth power-law evolution of the effective optical depth. This feature appears as a decrement with width $\Delta z \approx 0.1$ at $z \approx 3.2$. This feature is commonly attributed to He II reionization. Other studies do not detect such a feature, leaving doubt as to whether it arises from astrophysical sources or systematic errors.

To test for the presence of this He II feature, we use the mean transmission estimates as a function of redshift, and their covariances, from Section~\ref{sec:transmision}. As in previous studies, the effective optical depth is fit with a model of the form:
\begin{equation}
\tau_\mathrm{eff} = \tau_0 (1 + z)^\gamma + A \exp \left[\frac{-(z - z_\mathrm{cen})^2}{2 \sigma^2} \right];
\end{equation}
to the measured transmission. We searched for evidence of a Gaussian feature against a simple power-law model using the the BIC, with 40 data-points. There are two free parameters in the smooth model and five free parameters in the model with Gaussian departure.

\begin{figure}
\centering
\includegraphics[width=0.45\textwidth]{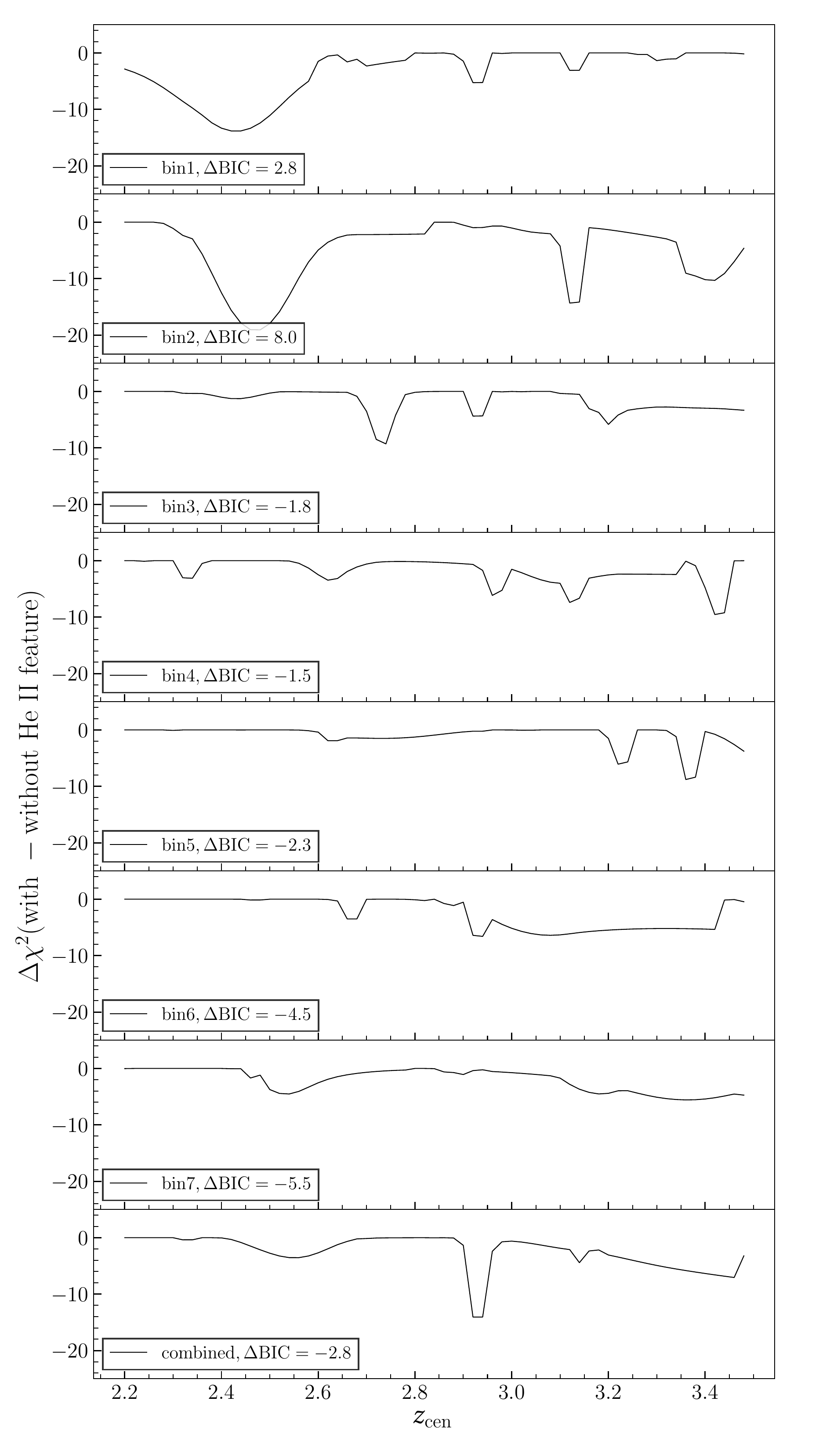}
\caption{$\Delta \chi^2$ surface as a function of the location of the Gaussian feature for the seven independent bins and for the combined analysis. The $\Delta \mathrm{BIC}$ values quoted reflect the difference between the five parameter model and the power-law model, at the redshift giving the lowest $\Delta \chi^2$.}
\label{fig:chi2_scan}
\end{figure}

% A complex model w{}ill give a smaller $\chi^2$, however BIC penalizes this by adding an extra term that depends on the number of free parameters of the model. A model with lower BIC is preferred.
Figure~\ref{fig:chi2_scan} shows the $\Delta \chi^2$ surface for each of the seven bins, as a function of the location of the He II Gaussian bump. This $\Delta \chi^2$ is reported as the difference in $\chi^2$ between a five-component model and a two-component model. The $\Delta \mathrm{BIC}$ values shown are computed at the best-fit redshift of the Gaussian feature. Bins 1 and 2 produce positive $\Delta$BIC estimates at a redshift $z_\mathrm{cen} \approx 2.4$. None of the other bins produce a positive $\Delta$BIC nor do they indicate a local minimum at this redshift. The bottom panel displays the joint model to the 280 data-points that span all seven bins. We do see a global minimum in the $\chi^2$ surface at $z_\mathrm{cen} \approx 2.92$; however, the negative value of the $\Delta$BIC suggests that no meaningful information is provided by the additional free parameters.

\section{Conclusions}\label{sec:conclusion}
This work has produced the tightest constraints to date on the redshift evolution of the mean effective optical depth due to Ly$\alpha$ absorption by neutral hydrogen. The evolution is fully-described by a powerlaw with no convincing evidence for He~II reionization in the redshift interval $2.0<z<3.5$. The measurement of the powerlaw exponents differ by 0.59 between the extreme values over seven independent analyses and are discrepant by 3.3$\sigma$ in the two measurements that show the largest tension. The final measurements on the optical depth parameters yield $\tau_0 = 0.00554 \pm 0.00064$ and $\gamma = 3.182 \pm 0.074$ after excluding a single deviant measurement and combining the results produced by the other six datasets.

One can compare the high-redshift results of this work to those at low redshifts based on Hubble Space Telescope (HST) observations of active galactic nuclei (AGN). \cite{Danforth:2016aa} measured the average Ly$\alpha$ flux decrement at redshifts $0 < z < 0.4$ using a sample of 82 medium resolution spectra of UV-bright AGN obtained from the Cosmic Origins Spectrograph. They report best-fit optical depth model with $\tau_0 = 0.014 \pm 0.001$ and $\gamma=2.2 \pm 0.2$. The shallower evolution is consistent with our model if we allow a break in the power-law at a redshift around $z = 1.6$. Future measurements of optical depth evolution over the redshift range $1 < z < 2$ will allow a direct test for such a break.

Systematic errors in the analysis appear to dominate the final measurement uncertainty. The uncertainty on the powerlaw exponent in the combined analysis is diluted from a statistical uncertainty of 1.2\% to 2.3\% when incorporating these systematic errors. Even after including these
systematic errors, the measurement produces a precision that is a factor of 1.6 better than
the previous measurement using the SDSS quasar sample \citep{Becker:2013aa}. The improvement is enabled by the larger sample of quasar spectra from the BOSS program. Beyond the increased sample size, the analysis is improved over previous work by the inclusion of covariances due to Large Scale
Structure and the division by quasar diversity. The former of these effects has been neglected in prior work but reduces the effective number of independent SDSS/BOSS wavelength bins by a factor of four. The latter allows a characterization of the source of systematic errors in the analysis.

Investigation of the systematic errors indicates that the measurements of spectral index used to normalize the Ly$\alpha$ forest continuum are subject to scatter and biased estimates. The algorithm in this analysis relies on a powerlaw fit to only 182 pixels over a wavelength range $1280 < \lambda_{\rm rf} < 1480$ \AA. Redshift-dependent variation in N V emission (1240~\AA), the O IV/Si IV emission line complex (1400~\AA), or C IV (1549~\AA) emission may lead to contamination in the region used to estimate the continuum and thus explain part of the bias.  Likewise, redshift dependence on the signal-to-noise ratio or some other affect of small statistical size of the continuum region may be biasing the powerlaw estimates. A future analysis may be able to mitigate these errors by taking a more comprehensive approach to continuum estimation and normalization. For example, archetype spectra can be identified based on a Set Cover Problem \citep[e.g.][]{Zhu:2016aa}, and a controlled sample of quasars can be identified around each archetype based on a $\chi^2$ nearest neighbor determination in the unabsorbed continuum at wavelengths longer than 1216 \AA. The spectral warping that we apply based on the simple continuum estimates could instead be performed using the entire unabsorbed wavelength region, thus producing a higher precision estimate that naturally incorporates emission line diversity.

Finally, we do not assess systematic errors due to flux calibration in this study, as such an analysis
is non-trivial given the lack of an independent broadband reference. The median residual when comparing eBOSS spectrophotometric fluxes to imaging fluxes in ($g, r, i, r - i$) have been shown to be ($-0.001, 0.004, -0.022, 0.032$) magnitudes, respectively \citep{Jensen:2016aa}. These systematic biases are at the level of the final precision in our analysis of the redshift evolving mean transmission. In addition, the lack of any strong deviation in the mean transmission from a powerlaw optical depth model provides evidence against significant spectroscopic calibration errors. The flux calibration errors would have to follow a powerlaw to avoid detection in that analysis.

The final sample from eBOSS exceeds 200,000 high-redshift quasar spectra. Such a sample can improve our results or extend the study by incorporating the Ly$\beta$ forest covering the restframe wavelength range $978 - 1014$ \AA\ \citep[e.g.][]{Irsic:2013aa}. Further improvement with the eBOSS spectra would require a reduction in the systematic errors to make use of the larger sample. An even larger sample of quasar spectra, with more stable flux calibration and higher $S/N$ ratio will be produced by the Dark Energy Spectroscopic Instrument \citep[DESI;][]{DESI-Collaboration:2016aa, DESI-Collaboration:2016ab} over the time period 2020--2025.  DESI will cover 14,000 square degrees and produce a sample of 700,000 quasar spectra at redshifts $z>2.1$. Observations of these quasars will have an effective exposure time roughly four times higher than that of eBOSS, thus leading to spectra that should be less susceptible to biases in continuum estimation arising from low $S/N$ ratio data. This DESI sample is designed for 1\% precision constraints on the Hubble parameter from BAO and will be very well suited for a new measurement of the mean transmission in the Ly$\alpha$ forest.

{\it Acknowledgments:}
\footnotesize
This paper represents an effort by both the SDSS-III and SDSS-IV collaborations. Funding for SDSS-III has been provided by the Alfred P. Sloan Foundation, the Participating Institutions, the National Science Foundation, and the U.S. Department of Energy Office of Science. The SDSS-III web site is http://www.sdss3.org/.

Funding for the Sloan Digital Sky Survey IV has been provided by the Alfred P. Sloan Foundation, the U.S. Department of Energy Office of Science, and the Participating Institutions. SDSS-IV acknowledges
support and resources from the Center for High-Performance Computing at
the University of Utah. The SDSS web site is www.sdss.org.

SDSS-IV is managed by the Astrophysical Research Consortium for the
Participating Institutions of the SDSS Collaboration including the
Brazilian Participation Group, the Carnegie Institution for Science,
Carnegie Mellon University, the Chilean Participation Group, the French Participation Group, Harvard-Smithsonian Center for Astrophysics,
Instituto de Astrof\'isica de Canarias, The Johns Hopkins University, Kavli Institute for the Physics and Mathematics of the Universe (IPMU) /
University of Tokyo, the Korean Participation Group, Lawrence Berkeley National Laboratory,
Leibniz Institut f\"ur Astrophysik Potsdam (AIP),
Max-Planck-Institut f\"ur Astronomie (MPIA Heidelberg),
Max-Planck-Institut f\"ur Astrophysik (MPA Garching),
Max-Planck-Institut f\"ur Extraterrestrische Physik (MPE),
National Astronomical Observatories of China, New Mexico State University,
New York University, University of Notre Dame,
Observat\'ario Nacional / MCTI, The Ohio State University,
Pennsylvania State University, Shanghai Astronomical Observatory,
United Kingdom Participation Group,
Universidad Nacional Aut\'onoma de M\'exico, University of Arizona,
University of Colorado Boulder, University of Oxford, University of Portsmouth,
University of Utah, University of Virginia, University of Washington, University of Wisconsin,
Vanderbilt University, and Yale University.

The work of VK, KD and HdMdB was supported in part by U.S. Department of Energy, Office of Science,
Office of High Energy Physics, under Award Number DESC0009959.

\newpage
\bibliographystyle{apj}
\bibliography{quasar_ms}

\end{document}